%% file: paper.tex
\documentclass[twocolumn, superscriptaddress]{revtex4-1} \usepackage{amsmath} \usepackage{booktabs}
\usepackage[utf8]{inputenc} \usepackage{xcolor} \usepackage{graphicx} \usepackage{hyperref}
\usepackage{orcidlink}
\usepackage{lineno}
\usepackage{acronym}

\newcommand{\tabref}[1]{Tab.~\ref{#1}}
\renewcommand{\eqref}[1]{Eq.~(\ref{#1})}

\newcommand{\figref}[1]{Fig.~\ref{#1}}
\newcommand{\secref}[1]{Sec.~\ref{#1}}

\begin{document}

\title{How Many Times Should We Matched Filter Gravitational Wave Data? A Comparison of GstLAL's Online and Offline Performance}

\input{authors.tex}

\date{\today}

\keywords{Suggested keywords} 

\begin{abstract}
\input{abstract.tex}
\end{abstract}

\maketitle

\input{acronyms.tex}
\input{macros.tex}

\input{introduction.tex}
\input{software_A.tex}
\input{software_B.tex}
\input{software_C.tex}

\input{methodology.tex}

\input{results_A.tex}
\input{results_B.tex}

\input{results_C.tex}
\input{results_D.tex}

\input{conclusion.tex}

\begin{acknowledgments}
\input{ack.tex}
\end{acknowledgments}


\clearpage

\bibliography{references}

\end{document}

%% file: authors.tex
\author{Prathamesh Joshi \orcidlink{0000-0002-4148-4932}}
\email{prathamesh.joshi@ligo.org}
\affiliation{Department of Physics, The Pennsylvania State University, University Park, PA 16802, USA}
\affiliation{Institute for Gravitation and the Cosmos, The Pennsylvania State University, University Park, PA 16802, USA}
\affiliation{School of Physics, Georgia Institute of Technology, Atlanta, GA 30332, USA}

\author{Wanting Niu \orcidlink{0000-0003-1470-532X}}
\email{wanting.niu@ligo.org}
\affiliation{Department of Physics, The Pennsylvania State University, University Park, PA 16802, USA}
\affiliation{Institute for Gravitation and the Cosmos, The Pennsylvania State University, University Park, PA 16802, USA}

\author{Chad Hanna}
\affiliation{Department of Physics, The Pennsylvania State University, University Park, PA 16802, USA}
\affiliation{Institute for Gravitation and the Cosmos, The Pennsylvania State University, University Park, PA 16802, USA}
\affiliation{Department of Astronomy and Astrophysics, The Pennsylvania State University, University Park, PA 16802, USA}
\affiliation{Institute for Computational and Data Sciences, The Pennsylvania State University, University Park, PA 16802, USA}

\author{Rachael Huxford}
\affiliation{Minnesota Supercomputing Institute, University of Minnesota, Minneapolis, MN 55455, USA}

\author{Divya Singh \orcidlink{0000-0001-9675-4584}}
\affiliation{Department of Physics, The Pennsylvania State University, University Park, PA 16802, USA}
\affiliation{Institute for Gravitation and the Cosmos, The Pennsylvania State University, University Park, PA 16802, USA}
\affiliation{Department of Physics, University of California, Berkeley, CA 94720, USA}

\author{Leo Tsukada  \orcidlink{0000-0003-0596-5648}}
\affiliation{Department of Physics, The Pennsylvania State University, University Park, PA 16802, USA}
\affiliation{Institute for Gravitation and the Cosmos, The Pennsylvania State University, University Park, PA 16802, USA}
\affiliation{Department of Physics and Astronomy, University of Nevada, Las Vegas, 4505 South Maryland Parkway, Las Vegas, NV 89154, USA}
\affiliation{Nevada Center for Astrophysics, University of Nevada, Las Vegas, NV 89154, USA}

\author{Shomik Adhicary \orcidlink{0009-0004-2101-5428}}
\affiliation{Department of Physics, The Pennsylvania State University, University Park, PA 16802, USA}
\affiliation{Institute for Gravitation and the Cosmos, The Pennsylvania State University, University Park, PA 16802, USA}

\author{Pratyusava Baral \orcidlink{0000-0001-6308-211X}}
\affiliation{Leonard E.\ Parker Center for Gravitation, Cosmology, and Astrophysics, University of Wisconsin-Milwaukee, Milwaukee, WI 53201, USA}

\author{Amanda Baylor \orcidlink{0000-0003-0918-0864}}
\affiliation{Leonard E.\ Parker Center for Gravitation, Cosmology, and Astrophysics, University of Wisconsin-Milwaukee, Milwaukee, WI 53201, USA}

\author{Kipp Cannon \orcidlink{0000-0003-4068-6572}}
\affiliation{RESCEU, The University of Tokyo, Tokyo, 113-0033, Japan}

\author{Sarah Caudill}
\affiliation{Department of Physics, University of Massachusetts, Dartmouth, MA 02747, USA}
\affiliation{Center for Scientific Computing and Data Science Research, University of Massachusetts, Dartmouth, MA 02747, USA}

\author{Michael W. Coughlin \orcidlink{0000-0002-8262-2924}}
\affiliation{School of Physics and Astronomy, University of Minnesota, Minneapolis, Minnesota 55455, USA}

\author{Bryce Cousins \orcidlink{0000-0002-7026-1340}}
\affiliation{Department of Physics, University of Illinois, Urbana, IL 61801 USA}
\affiliation{Department of Physics, The Pennsylvania State University, University Park, PA 16802, USA}
\affiliation{Institute for Gravitation and the Cosmos, The Pennsylvania State University, University Park, PA 16802, USA}

\author{Jolien D. E. Creighton \orcidlink{0000-0003-3600-2406}}
\affiliation{Leonard E.\ Parker Center for Gravitation, Cosmology, and Astrophysics, University of Wisconsin-Milwaukee, Milwaukee, WI 53201, USA}

\author{Becca Ewing}
\affiliation{Department of Physics, The Pennsylvania State University, University Park, PA 16802, USA}
\affiliation{Institute for Gravitation and the Cosmos, The Pennsylvania State University, University Park, PA 16802, USA}

\author{Heather Fong}
\affiliation{Department of Physics and Astronomy, University of British Columbia, Vancouver, BC, V6T 1Z4, Canada}
\affiliation{RESCEU, The University of Tokyo, Tokyo, 113-0033, Japan}
\affiliation{Graduate School of Science, The University of Tokyo, Tokyo 113-0033, Japan}

\author{Richard N. George \orcidlink{0000-0002-7797-7683}}
\affiliation{Center for Gravitational Physics, University of Texas at Austin, Austin, TX 78712, USA}

\author{Shaon Ghosh}
\affiliation{Montclair State University, 1 Normal Ave, Montclair, NJ 07042}

\author{Patrick Godwin \orcidlink{0000-0002-7489-4751}}
\affiliation{LIGO Laboratory, California Institute of Technology, MS 100-36, Pasadena, California 91125, USA}
\affiliation{Department of Physics, The Pennsylvania State University, University Park, PA 16802, USA}
\affiliation{Institute for Gravitation and the Cosmos, The Pennsylvania State University, University Park, PA 16802, USA}

\author{Reiko Harada}
\affiliation{RESCEU, The University of Tokyo, Tokyo, 113-0033, Japan}
\affiliation{Graduate School of Science, The University of Tokyo, Tokyo 113-0033, Japan}

\author{Yun-Jing Huang \orcidlink{0000-0002-2952-8429}}
\affiliation{Department of Physics, The Pennsylvania State University, University Park, PA 16802, USA}
\affiliation{Institute for Gravitation and the Cosmos, The Pennsylvania State University, University Park, PA 16802, USA}

\author{James Kennington \orcidlink{0000-0002-6899-3833}}
\affiliation{Department of Physics, The Pennsylvania State University, University Park, PA 16802, USA}
\affiliation{Institute for Gravitation and the Cosmos, The Pennsylvania State University, University Park, PA 16802, USA}

\author{Soichiro Kuwahara}
\affiliation{RESCEU, The University of Tokyo, Tokyo, 113-0033, Japan}
\affiliation{Graduate School of Science, The University of Tokyo, Tokyo 113-0033, Japan}

\author{Alvin K. Y. Li \orcidlink{0000-0001-6728-6523}}
\affiliation{RESCEU, The University of Tokyo, Tokyo, 113-0033, Japan}
\affiliation{Graduate School of Science, The University of Tokyo, Tokyo 113-0033, Japan}

\author{Ryan Magee \orcidlink{0000-0001-9769-531X}}
\affiliation{LIGO Laboratory, California Institute of Technology, Pasadena, CA 91125, USA}

\author{Duncan Meacher \orcidlink{0000-0001-5882-0368}}
\affiliation{Leonard E.\ Parker Center for Gravitation, Cosmology, and Astrophysics, University of Wisconsin-Milwaukee, Milwaukee, WI 53201, USA}

\author{Cody Messick \orcidlink{0000-0002-8230-3309}}
\affiliation{Leonard E.\ Parker Center for Gravitation, Cosmology, and Astrophysics, University of Wisconsin-Milwaukee, Milwaukee, WI 53201, USA}

\author{Soichiro Morisaki \orcidlink{0000-0002-8445-6747}}
\affiliation{Institute for Cosmic Ray Research, The University of Tokyo, 5-1-5 Kashiwanoha, Kashiwa, Chiba 277-8582, Japan}

\author{Debnandini Mukherjee  \orcidlink{0000-0001-7335-9418}}
\affiliation{NASA Marshall Space Flight Center, Huntsville, AL 35811, USA}
\affiliation{Center for Space Plasma and Aeronomic Research, University of Alabama in Huntsville, Huntsville, AL 35899, USA}

\author{Alexander Pace \orcidlink{0009-0003-4044-0334}}
\affiliation{Department of Physics, The Pennsylvania State University, University Park, PA 16802, USA}
\affiliation{Institute for Gravitation and the Cosmos, The Pennsylvania State University, University Park, PA 16802, USA}

\author{Cort Posnansky \orcidlink{0009-0009-7137-9795}}
\affiliation{Department of Physics, The Pennsylvania State University, University Park, PA 16802, USA}
\affiliation{Institute for Gravitation and the Cosmos, The Pennsylvania State University, University Park, PA 16802, USA}

\author{Anarya Ray \orcidlink{0000-0002-7322-4748}}
\affiliation{Leonard E.\ Parker Center for Gravitation, Cosmology, and Astrophysics, University of Wisconsin-Milwaukee, Milwaukee, WI 53201, USA}
\affiliation{Center of Interdisciplinary Education and Research in Astrophysics, Northwestern University, IL 60201, USA}

\author{Surabhi Sachdev \orcidlink{0000-0002-0525-2317}}
\affiliation{School of Physics, Georgia Institute of Technology, Atlanta, GA 30332, USA}
\affiliation{Leonard E.\ Parker Center for Gravitation, Cosmology, and Astrophysics, University of Wisconsin-Milwaukee, Milwaukee, WI 53201, USA}

\author{Shio Sakon \orcidlink{0000-0002-5861-3024}}
\affiliation{Department of Physics, The Pennsylvania State University, University Park, PA 16802, USA}
\affiliation{Institute for Gravitation and the Cosmos, The Pennsylvania State University, University Park, PA 16802, USA}

\author{Urja Shah \orcidlink{0000-0001-8249-7425}}
\affiliation{School of Physics, Georgia Institute of Technology, Atlanta, GA 30332, USA}

\author{Ron Tapia}
\affiliation{Department of Physics, The Pennsylvania State University, University Park, PA 16802, USA}
\affiliation{Institute for Computational and Data Sciences, The Pennsylvania State University, University Park, PA 16802, USA}

\author{Koh Ueno \orcidlink{0000-0003-3227-6055}}
\affiliation{RESCEU, The University of Tokyo, Tokyo, 113-0033, Japan}

\author{Aaron Viets \orcidlink{0000-0002-4241-1428}}
\affiliation{Concordia University Wisconsin, Mequon, WI 53097, USA}

\author{Leslie Wade}
\affiliation{Department of Physics, Hayes Hall, Kenyon College, Gambier, Ohio 43022, USA}

\author{Madeline Wade \orcidlink{0000-0002-5703-4469}}
\affiliation{Department of Physics, Hayes Hall, Kenyon College, Gambier, Ohio 43022, USA}

\author{Zach Yarbrough \orcidlink{0000-0002-9825-1136}}
\affiliation{Department of Physics and Astronomy, Louisiana State University, Baton Rouge, LA 70803, USA}

\author{Noah Zhang \orcidlink{0009-0003-3361-5538}}
\affiliation{School of Physics, Georgia Institute of Technology, Atlanta, GA 30332, USA}

%% file: abstract.tex
Searches for gravitational waves from compact binary coalescences employ a process called
matched filtering, in which gravitational wave strain data is cross-correlated against a
bank of waveform templates. Data from every observing run of the LIGO, Virgo, and KAGRA
collaboration is typically analyzed in this way twice, first in a low-latency mode in which
gravitational wave candidates are identified in near-real time, and later in a high-latency
mode. Such high-latency analyses have traditionally been considered more sensitive, since
background data from the full observing run is available for assigning significance to 
all candidates, as well as more robust, since they do not need to worry about keeping up with
live data. In this work, we present a novel technique to use the matched filtering data products from a low-latency
analysis and re-process them by assigning significances in a high-latency way, effectively removing the need to perform matched filtering a
second time. To demonstrate the efficacy of our method, we analyze 38 days of LIGO and Virgo data from the third
observing run (O3) using the GstLAL pipeline, and show that our method is as sensitive and reliable as a
traditional high-latency analysis. Since matched filtering represents the vast majority of
computing time for a traditional analysis, our method greatly reduces the time and computational burden required to 
produce the same results as a traditional high-latency analysis. Consequently, it has already been adopted by
GstLAL for the fourth observing run (O4) of the LIGO, Virgo, and KAGRA collaboration.

%% file: acronyms.tex
\acrodef{LSC}[LSC]{LIGO Scientific Collaboration}
\acrodef{LVC}[LVC]{LIGO Scientific and Virgo Collaboration}
\acrodef{LVK}[LVK]{LIGO Scientific, Virgo and KAGRA Collaboration}
\acrodef{aLIGO}{Advanced Laser Interferometer Gravitational-Wave Observatory}
\acrodef{aVirgo}{Advanced Virgo}
\acrodef{LIGO}[LIGO]{Laser Interferometer Gravitational-Wave Observatory}
\acrodef{IFO}[IFO]{interferometer}
\acrodef{LHO}[LHO]{LIGO-Hanford}
\acrodef{LLO}[LLO]{LIGO-Livingston}
\acrodef{O2}[O2]{second observing run}
\acrodef{O1}[O1]{first observing run}
\acrodef{O3}[O3]{third observing run}
\acrodef{O3a}[O3a]{first half of the third observing run}
\acrodef{O3b}[O3b]{second half of the third observing run}
\acrodef{O4a}[O4a]{first part of the fourth observing run}
\acrodef{O4}[O4]{fourth observing run}

\acrodef{SSM}[SSM]{subsolar-mass}
\acrodef{BH}[BH]{black hole}
\acrodef{BBH}[BBH]{binary black hole}
\acrodef{BNS}[BNS]{binary neutron star}
\acrodef{IMBH}[IMBH]{intermediate-mass black hole}
\acrodef{NS}[NS]{neutron star}
\acrodef{BHNS}[BHNS]{black hole--neutron star binaries}
\acrodef{NSBH}[NSBH]{neutron star--black hole binary}
\acrodef{PBH}[PBH]{primordial black hole binaries}
\acrodef{CBC}[CBC]{compact binary coalescence}
\acrodef{GW}[GW]{gravitational wave}
\acrodef{GWH}[GW]{gravitational-wave}
\acrodef{DBH}[DBH]{dispasstive black hole binaries}

\acrodef{SNR}[SNR]{signal-to-noise ratio}
\acrodef{FAR}[FAR]{false alarm rate}
\acrodef{PSD}[PSD]{power spectral density}
\acrodef{GR}[GR]{general relativity}
\acrodef{NR}[NR]{numerical relativity}
\acrodef{PN}[PN]{post-Newtonian}
\acrodef{EOB}[EOB]{effective-one-body}
\acrodef{ROM}[ROM]{reduced-order model}
\acrodef{IMR}[IMR]{inspiral--merger--ringdown}
\acrodef{EOS}[EoS]{equation of state}
\acrodef{FF}[FF]{fitting factor}
\acrodef{FT}[FT]{Fourier Transform}

\acrodef{LAL}[LAL]{LIGO Algorithm Library}
\acrodef{GWTC}[GWTC]{Gravitational Wave Transient Catalog}

\newcommand{\PN}[0]{\ac{PN}\xspace}
\newcommand{\BBH}[0]{\ac{BBH}\xspace}
\newcommand{\BNS}[0]{\ac{BNS}\xspace}
\newcommand{\BH}[0]{\ac{BH}\xspace}
\newcommand{\NR}[0]{\ac{NR}\xspace}
\newcommand{\GW}[0]{\ac{GW}\xspace}
\newcommand{\SNR}[0]{\ac{SNR}\xspace}
\newcommand{\SSM}[0]{\ac{SSM}\xspace}
\newcommand{\aLIGO}[0]{\ac{aLIGO}\xspace}
\newcommand{\PSD}[0]{\ac{PSD}\xspace}
\newcommand{\GR}[0]{\ac{GR}\xspace}
\newcommand{\EOS}[0]{\ac{EOS}\xspace}
\newcommand{\LVC}[0]{\ac{LVC}\xspace}


\newcommand{\GSTLAL}{GstLAL\xspace}
\newcommand{\IMRPHENOMD}{IMRPhenomD\xspace}
\newcommand{\MANIFOLD}{{\fontfamily{qcr}\selectfont manifold}\xspace}
\newcommand{\SBANK}{{\fontfamily{qcr}\selectfont SBank}\xspace}

%% file: macros.tex
\newcommand\hmm[1]{\ifnum\ifhmode\spacefactor\else2000\fi>1500 \uppercase{#1}\else#1\fi}

\newcommand{\IGWNALERT}{\texttt{igwn-alert}\xspace}

\newcommand{\MDCSTART}{5 Jan. 2020 15:59:42}
\newcommand{\MDCEND}{14 Feb. 2020 15:59:42}

\newcommand{\TOTALTEMPLATES}{\ensuremath{1.8 \times 10^6}}
\newcommand{\CHECKERBOARDTEMPLATES}{\ensuremath{9 \times 10^5}}
\newcommand{\NUMSVDBANKS}{\ensuremath{\sim 1000}}
\newcommand{\TEMPLATESPERSUBBANK}{\ensuremath{\sim 500}}
\newcommand{\NUMSUBBANKSPERSVD}{\ensuremath{2}}
\newcommand{\SVDTOLERANCE}{\ensuremath{99.999\%}}
\newcommand{\PSDFFTLENGTH}{\ensuremath{4~\mathrm{seconds}}}
\newcommand{\FRAMELENGTH}{\ensuremath{1~\mathrm{second}}}
\newcommand{\BUFFERBLOCKSIZE}{\ensuremath{4096~\mathrm{bytes}}}
\newcommand{\FIRSTRIDE}{\ensuremath{0.25~\mathrm{seconds}}}
\newcommand{\TRIGGERSNRTHRESHOLD}{\ensuremath{4.0}}
\newcommand{\COINCTHRESHOLD}{\ensuremath{0.005~\mathrm{seconds}}}
\newcommand{\HTGATETHRESHOLDMIN}{\ensuremath{15.0}}
\newcommand{\HTGATETHRESHOLDMAX}{\ensuremath{100.0}}
\newcommand{\HTGATEMCHIRPMIN}{\ensuremath{0.8}}
\newcommand{\HTGATEMCHIRPMAX}{\ensuremath{45.0}}
\newcommand{\HTGATEMIN}{\ensuremath{\sim 15}}
\newcommand{\HTGATEMAX}{\ensuremath{\sim 325}}
\newcommand{\LRSNAPSHOT}{\ensuremath{4}}
\newcommand{\LRCOMPRESSION}{\ensuremath{0.003}}
\newcommand{\FARTRIALSFACTOR}{\ensuremath{2}}
\newcommand{\UPLOADCADENCE}{\ensuremath{4}}
\newcommand{\UPLOADCADENCEMDCTWELVE}{\ensuremath{2}}
\newcommand{\UPLOADDT}{\ensuremath{0.2}}
\newcommand{\SINGLESPENALTYMDCELEVEN}{\ensuremath{12}}
\newcommand{\SINGLESPENALTYOFOUR}{\ensuremath{13}}
\newcommand{\XISQMISMATCHRANGE}{\ensuremath{0.1-10\%}}

\newcommand{\TESTSUITECOINCWINDOW}{\ensuremath{\pm 1}}
\newcommand{\INJSNRFLOW}{\ensuremath{10.0}}
\newcommand{\INJSNRFHI}{\ensuremath{1600.0}}

\newcommand{\VT}{\ensuremath{\langle VT \rangle}}
\newcommand{\SPINZ}{\ensuremath{s_{i,z}}}
\newcommand{\CHIP}{\ensuremath{\chi_p}}
\newcommand{\MCHIRP}{\ensuremath{\mathcal{M}_c}\xspace}
\newcommand{\CHIEFF}{\ensuremath{\chi_{\mathrm{eff}}}}
\newcommand{\PASTRO}{\ensuremath{p(\mathrm{astro})}}
\newcommand{\MSUN}{\ensuremath{M_{\odot}}}
\newcommand{\TEND}{\ensuremath{t_{\mathrm{end}}}}

\newcommand{\TOTALINJECTIONS}{$5\times10^4$}
\newcommand{\BNSMAXZ}{\ensuremath{0.15}}
\newcommand{\NSBHMAXZ}{\ensuremath{0.25}}
\newcommand{\BBHMAXZ}{\ensuremath{1.9}}
\newcommand{\MDCDURATION}{\ensuremath{3.456\times10^6}}
\newcommand{\INJECTIONSPACING}{\ensuremath{\sim40}}

\newcommand{\PASTROTHRESHOLD}{\ensuremath{0.50}}

\newcommand{\DECISIVESNRTHRESH}{\ensuremath{8.0}}
\newcommand{\NETWORKSNRTHRESH}{\ensuremath{10.0}}

\newcommand{\ALLABOVEDECSNRTHRESH}{\ensuremath{1522}}
\newcommand{\ALLINBANKABOVEDECSNRTHRESH}{\ensuremath{1457}}
\newcommand{\BBHINBANKABOVEDECSNRTHRESH}{\ensuremath{597}}
\newcommand{\BNSINBANKABOVEDECSNRTHRESH}{\ensuremath{482}}
\newcommand{\NSBHINBANKABOVEDECSNRTHRESH}{\ensuremath{378}}

\newcommand{\HIGHFARTHRESH}{$1$ per hour}
\newcommand{\LOWFARTHRESH}{$2$ per day}
\newcommand{\ONEPERHOUR}{\ensuremath{2.78\times10^{-4}~\mathrm{Hz}}}
\newcommand{\TWOPERDAY}{\ensuremath{2.31\times10^{-5}~\mathrm{Hz}}}
\newcommand{\ONEPERMONTH}{\ensuremath{3.85\times10^{-7}~\mathrm{Hz}}}
\newcommand{\TWOPERYEAR}{\ensuremath{3.16\times10^{-8}~\mathrm{Hz}}}

\newcommand{\ALLINBANKEFFICIENCY}[1]{%
	\IfEqCase{#1}{%
		{ONEPERHOUR}{\ensuremath{0.87}}%
		{TWOPERDAY}{\ensuremath{0.84}}%
		{ONEPERMONTH}{\ensuremath{0.78}}%
		{TWOPERYEAR}{\ensuremath{0.74}}%
	}[\PackageError{ALLINBANKEFFICIENCY}{Undefined option: #1}{}]
}%

\newcommand{\BBHINBANKEFFICIENCY}[1]{%
	\IfEqCase{#1}{%
		{ONEPERHOUR}{\ensuremath{0.87}}%
		{TWOPERDAY}{\ensuremath{0.84}}%
		{ONEPERMONTH}{\ensuremath{0.77}}%
		{TWOPERYEAR}{\ensuremath{0.71}}%
	}[\PackageError{BBHINBANKEFFICIENCY}{Undefined option: #1}{}]
}%

\newcommand{\BNSINBANKEFFICIENCY}[1]{%
	\IfEqCase{#1}{%
		{ONEPERHOUR}{\ensuremath{0.95}}%
		{TWOPERDAY}{\ensuremath{0.95}}%
		{ONEPERMONTH}{\ensuremath{0.89}}%
		{TWOPERYEAR}{\ensuremath{0.86}}%
	}[\PackageError{BNSINBANKEFFICIENCY}{Undefined option: #1}{}]
}%

\newcommand{\NSBHINBANKEFFICIENCY}[1]{%
	\IfEqCase{#1}{%
		{ONEPERHOUR}{\ensuremath{0.77}}%
		{TWOPERDAY}{\ensuremath{0.71}}%
		{ONEPERMONTH}{\ensuremath{0.65}}%
		{TWOPERYEAR}{\ensuremath{0.62}}%
	}[\PackageError{NSBHINBANKEFFICIENCY}{Undefined option: #1}{}]
}%

\newcommand{\ALLABOVEDECSNRTHRESHMDCTWELVE}{\ensuremath{653}}
\newcommand{\ALLINBANKABOVEDECSNRTHRESHMDCTWELVE}{\ensuremath{621}}
\newcommand{\BBHINBANKABOVEDECSNRTHRESHMDCTWELVE}{\ensuremath{243}}
\newcommand{\BNSINBANKABOVEDECSNRTHRESHMDCTWELVE}{\ensuremath{209}}
\newcommand{\NSBHINBANKABOVEDECSNRTHRESHMDCTWELVE}{\ensuremath{169}}

\newcommand{\ALLINBANKEFFICIENCYMDCTWELVE}[1]{%
	\IfEqCase{#1}{%
		{ONEPERHOUR}{\ensuremath{0.88}}%
		{TWOPERDAY}{\ensuremath{0.86}}%
		{ONEPERMONTH}{\ensuremath{0.83}}%
		{TWOPERYEAR}{\ensuremath{0.81}}%
	}[\PackageError{ALLINBANKEFFICIENCYMDCTWELVE}{Undefined option: #1}{}]
}%

\newcommand{\BBHINBANKEFFICIENCYMDCTWELVE}[1]{%
	\IfEqCase{#1}{%
		{ONEPERHOUR}{\ensuremath{0.92}}%
		{TWOPERDAY}{\ensuremath{0.90}}%
		{ONEPERMONTH}{\ensuremath{0.87}}%
		{TWOPERYEAR}{\ensuremath{0.86}}%
	}[\PackageError{BBHINBANKEFFICIENCYMDCTWELVE}{Undefined option: #1}{}]
}%

\newcommand{\BNSINBANKEFFICIENCYMDCTWELVE}[1]{%
	\IfEqCase{#1}{%
		{ONEPERHOUR}{\ensuremath{0.95}}%
		{TWOPERDAY}{\ensuremath{0.93}}%
		{ONEPERMONTH}{\ensuremath{0.91}}%
		{TWOPERYEAR}{\ensuremath{0.88}}%
	}[\PackageError{BNSINBANKEFFICIENCYMDCTWELVE}{Undefined option: #1}{}]
}%

\newcommand{\NSBHINBANKEFFICIENCYMDCTWELVE}[1]{%
	\IfEqCase{#1}{%
		{ONEPERHOUR}{\ensuremath{0.74}}%
		{TWOPERDAY}{\ensuremath{0.72}}%
		{ONEPERMONTH}{\ensuremath{0.68}}%
		{TWOPERYEAR}{\ensuremath{0.66}}%
	}[\PackageError{NSBHINBANKEFFICIENCYMDCTWELVE}{Undefined option: #1}{}]
}%

\newcommand{\MEAN}[1]{%
	\IfEqCase{#1}{%
		{MASSRATIO}{\ensuremath{1.39}}%
		{MCHIRP}{\ensuremath{0.15}}%
		{SPIN1Z}{\ensuremath{7.27}}%
		{SPIN2Z}{\ensuremath{2.81}}%
		{CHIEFF}{\ensuremath{5.77}}%
		{ENDTIME}{\ensuremath{6.23}}%
	}[\PackageError{MEAN}{Undefined option: #1}{}]
}%

\newcommand{\STDEV}[1]{%
	\IfEqCase{#1}{%
		{MASSRATIO}{\ensuremath{2.86}}%
		{MCHIRP}{\ensuremath{0.45}}%
		{SPIN1Z}{\ensuremath{285}}%
		{SPIN2Z}{\ensuremath{183}}%
		{CHIEFF}{\ensuremath{252}}%
		{ENDTIME}{\ensuremath{30.22}}%
	}[\PackageError{STDEV}{Undefined option: #1}{}]
}%

\newcommand{\BNSMCHIRPMEAN}{\ensuremath{2.06\times10^{-4}}}
\newcommand{\BNSMCHIRPSTDEV}{\ensuremath{8.33\times10^{-4}}}

\newcommand{\NSBHMCHIRPMEAN}{\ensuremath{-2.14\times10^{-4}}}
\newcommand{\NSBHMCHIRPSTDEV}{\ensuremath{6.26\times10^{-3}}}

\newcommand{\BBHMCHIRPMEAN}{\ensuremath{1.54\times10^{-1}}}
\newcommand{\BBHMCHIRPSTDEV}{\ensuremath{4.53\times10^{-1}}}

\newcommand{\BNSENDTIMEMEAN}{\ensuremath{-0.90}}
\newcommand{\BNSENDTIMESTDEV}{\ensuremath{18.0}}

\newcommand{\NSBHENDTIMEMEAN}{\ensuremath{18.7}}
\newcommand{\NSBHENDTIMESTDEV}{\ensuremath{59.3}}

\newcommand{\BBHENDTIMEMEAN}{\ensuremath{6.03}}
\newcommand{\BBHENDTIMESTDEV}{\ensuremath{11.3}}

\newcommand{\QFIFTY}[1]{%
	\IfEqCase{#1}{%
		{MASSRATIO}{\ensuremath{0.45}}%
		{MCHIRP}{\ensuremath{0.007}}%
		{SPIN1Z}{\ensuremath{1.24}}%
		{SPIN2Z}{\ensuremath{1.72}}%
		{CHIEFF}{\ensuremath{1.34}}%
		{ENDTIME}{\ensuremath{3.8}}%
	}[\PackageError{QFIFTY}{Undefined option: #1}{}]
}%

\newcommand{\QSEVENTYFIVE}[1]{%
	\IfEqCase{#1}{%
		{MASSRATIO}{\ensuremath{1.67}}%
		{MCHIRP}{\ensuremath{0.33}}%
		{SPIN1Z}{\ensuremath{3.82}}%
		{SPIN2Z}{\ensuremath{5.33}}%
		{CHIEFF}{\ensuremath{3.71}}%
		{ENDTIME}{\ensuremath{9.78}}%
	}[\PackageError{QSEVENTYFIVE}{Undefined option: #1}{}]
}%

\newcommand{\QNINETY}[1]{%
	\IfEqCase{#1}{%
		{MASSRATIO}{\ensuremath{4.97}}%
		{MCHIRP}{\ensuremath{0.73}}%
		{SPIN1Z}{\ensuremath{13.8}}%
		{SPIN2Z}{\ensuremath{17.5}}%
		{CHIEFF}{\ensuremath{10.8}}%
		{ENDTIME}{\ensuremath{25.6}}%
	}[\PackageError{QNINETY}{Undefined option: #1}{}]
}%

\newcommand{\GPCYRS}{\ensuremath{\mathrm{Gpc}^3\mathrm{yrs}}}
\newcommand{\INJECTEDVT}[1]{%
	\IfEqCase{#1}{%
		{BNS}{\ensuremath{1.08\times10^{-1}}}%
		{NSBH}{\ensuremath{4.34\times10^{-1}}}%
		{BBH}{\ensuremath{29.1}}%
	}[\PackageError{INJECTEDVT}{Undefined option: #1}{}]
}%

\newcommand{\VTTWOPERDAY}[1]{%
	\IfEqCase{#1}{%
		{BNS}{\ensuremath{3.49\times10^{-4}}}%
		{NSBH}{\ensuremath{8.08\times10^{-4}}}%
		{BBH}{\ensuremath{1.23\times10^{-1}}}%
	}[\PackageError{VTTWOPERDAY}{Undefined option: #1}{}]
}%

\newcommand{\VTNETSNR}[1]{%
	\IfEqCase{#1}{%
		{BNS}{\ensuremath{4.41\times10^{-4}}}%
		{NSBH}{\ensuremath{1.59\times10^{-3}}}%
		{BBH}{\ensuremath{1.52\times10^{-1}}}%
	}[\PackageError{VTNETSNR}{Undefined option: #1}{}]
}%

\newcommand{\VTDECSNR}[1]{%
	\IfEqCase{#1}{%
		{BNS}{\ensuremath{1.47\times10^{-4}}}%
		{NSBH}{\ensuremath{4.98\times10^{-3}}}%
		{BBH}{\ensuremath{5.46\times10^{-2}}}%
	}[\PackageError{VTDECSNR}{Undefined option: #1}{}]
}%


\newcommand{\SEARCHEDAREAQFIFTY}[1]{%
	\IfEqCase{#1}{%
		{ALL}{\ensuremath{271}}%
		{TRIPLE}{\ensuremath{31.9}}%
		{DOUBLE}{\ensuremath{301}}%
		{SINGLE}{\ensuremath{3150}}%
	}[\PackageError{SEARCHEDAREAQFIFTY}{Undefined option: #1}{}]
}%

\newcommand{\SEARCHEDAREAQSEVENTYFIVE}[1]{%
	\IfEqCase{#1}{%
		{ALL}{\ensuremath{1080}}%
		{TRIPLE}{\ensuremath{140}}%
		{DOUBLE}{\ensuremath{893}}%
		{SINGLE}{\ensuremath{10,400}}%
	}[\PackageError{SEARCHEDAREAQSEVENTYFIVE}{Undefined option: #1}{}]
}%

\newcommand{\SEARCHEDAREAQNINETY}[1]{%
	\IfEqCase{#1}{%
		{ALL}{\ensuremath{3910}}%
		{TRIPLE}{\ensuremath{357}}%
		{DOUBLE}{\ensuremath{2470}}%
		{SINGLE}{\ensuremath{18,400}}%
	}[\PackageError{SEARCHEDAREAQNINETY}{Undefined option: #1}{}]
}%

\newcommand{\SEARCHEDPROBQFIFTY}[1]{%
	\IfEqCase{#1}{%
		{ALL}{\ensuremath{0.53}}%
		{TRIPLE}{\ensuremath{0.58}}%
		{DOUBLE}{\ensuremath{0.52}}%
		{SINGLE}{\ensuremath{0.59}}%
	}[\PackageError{SEARCHEDPROBQFIFTY}{Undefined option: #1}{}]
}%

\newcommand{\SEARCHEDPROBQSEVENTYFIVE}[1]{%
	\IfEqCase{#1}{%
		{ALL}{\ensuremath{0.79}}%
		{TRIPLE}{\ensuremath{0.84}}%
		{DOUBLE}{\ensuremath{0.77}}%
		{SINGLE}{\ensuremath{0.78}}%
	}[\PackageError{SEARCHEDPROBQSEVENTYFIVE}{Undefined option: #1}{}]
}%

\newcommand{\SEARCHEDPROBQNINETY}[1]{%
	\IfEqCase{#1}{%
		{ALL}{\ensuremath{0.93}}%
		{TRIPLE}{\ensuremath{0.96}}%
		{DOUBLE}{\ensuremath{0.92}}%
		{SINGLE}{\ensuremath{0.92}}%
	}[\PackageError{SEARCHEDPROBQNINETY}{Undefined option: #1}{}]
}%

\newcommand{\BNSTOBNS}{\ensuremath{90.3\%}}
\newcommand{\BNSTONSBH}{\ensuremath{9.7\%}}

\newcommand{\NSBHTONSBH}{\ensuremath{64.1\%}}
\newcommand{\NSBHTOBBH}{\ensuremath{33.8\%}}
\newcommand{\NSBHTOBNS}{\ensuremath{2.10\%}}

\newcommand{\BBHTOBBH}{\ensuremath{100\%}}

\newcommand{\TERRTOTERR}{\ensuremath{2.60\%}}
\newcommand{\TERRTOBBH}{\ensuremath{68.8\%}}
\newcommand{\TERRTONSBH}{\ensuremath{19.5\%}}
\newcommand{\TERRTOBNS}{\ensuremath{9.10\%}}

\newcommand{\BNSTOBNSMDCTWELVE}{\ensuremath{79.8\%}}
\newcommand{\BNSTONSBHMDCTWELVE}{\ensuremath{20.2\%}}

\newcommand{\NSBHTONSBHMDCTWELVE}{\ensuremath{92.1\%}}
\newcommand{\NSBHTOBBHMDCTWELVE}{\ensuremath{6.83\%}}
\newcommand{\NSBHTOBNSMDCTWELVE}{\ensuremath{1.02\%}}

\newcommand{\BBHTOBBHMDCTWELVE}{\ensuremath{99.5\%}}
\newcommand{\BBHTONSBHMDCTWELVE}{\ensuremath{0.05\%}}

\newcommand{\TERRTOBBHMDCTWELVE}{\ensuremath{76.2\%}}
\newcommand{\TERRTONSBHMDCTWELVE}{\ensuremath{23.8\%}}

\newcommand{\OTHREEOPA}{\ensuremath{1.2}} 

\newcommand{\MDCGWIFOS}[1]{%
	\IfEqCase{#1}{%
		{GW200112}{L1}%
		{GW200115}{H1L1}%
		{GW200128}{H1L1}%
		{GW200129}{H1L1V1}%
		{GW200202}{H1L1}%
		{GW200208q}{H1L1}%
		{GW200208am}{H1L1}%
		{GW200209}{H1L1}%
		{GW200210}{H1L1}%
	}[\PackageError{MDCGWIFOS}{Undefined option: #1}{}]
}%

\newcommand{\MDCGWSNR}[1]{%
	\IfEqCase{#1}{%
		{GW200112}{\ensuremath{18.46}}%
		{GW200115}{\ensuremath{11.48}}%
		{GW200128}{\ensuremath{9.98}}%
		{GW200129}{\ensuremath{26.30}}%
		{GW200202}{\ensuremath{11.09}}%
		{GW200208q}{\ensuremath{10.56}}%
		{GW200208am}{\ensuremath{8.00}}%
		{GW200209}{\ensuremath{9.96}}%
		{GW200210}{\ensuremath{9.28}}%
	}[\PackageError{MDCGWSNR}{Undefined option: #1}{}]
}%

\newcommand{\MDCGWFAR}[1]{%
	\IfEqCase{#1}{%
		{GW200112}{\ensuremath{1.01\times10^{-7}}}%
		{GW200115}{\ensuremath{2.55\times10^{-4}}}%
		{GW200128}{\ensuremath{1.44\times10^{-4}}}%
		{GW200129}{\ensuremath{1.78\times10^{-17}}}%
		{GW200202}{\ensuremath{1.69\times10^{-2}}}%
		{GW200208q}{\ensuremath{4.92\times10^{-5}}}%
		{GW200208am}{\ensuremath{2.02\times10^{3}}}%
		{GW200209}{\ensuremath{1.20}}%
		{GW200210}{\ensuremath{3.64\times10^{3}}}%
	}[\PackageError{MDCGWFAR}{Undefined option: #1}{}]
}%

\newcommand{\MDCGWPASTRO}[1]{%
	\IfEqCase{#1}{%
		{GW200112}{\ensuremath{>0.99}}%
		{GW200115}{\ensuremath{>0.99}}%
		{GW200128}{\ensuremath{>0.99}}%
		{GW200129}{\ensuremath{>0.99}}%
		{GW200202}{\ensuremath{>0.99}}%
		{GW200208q}{\ensuremath{>0.99}}%
		{GW200208am}{\ensuremath{0.48}}%
		{GW200209}{\ensuremath{>0.99}}%
		{GW200210}{0.27}%
	}[\PackageError{MDCGWPASTRO}{Undefined option: #1}{}]
}%

\newcommand{\MDCGWMCHIRP}[1]{%
	\IfEqCase{#1}{%
		{GW200112}{\ensuremath{33.37~M_{\odot}}}%
		{GW200115}{\ensuremath{2.58~M_{\odot}}}%
		{GW200128}{\ensuremath{50.74~M_{\odot}}}%
		{GW200129}{\ensuremath{30.66~M_{\odot}}}%
		{GW200202}{\ensuremath{8.15~M_{\odot}}}%
		{GW200208q}{\ensuremath{34.50~M_{\odot}}}%
		{GW200208am}{\ensuremath{66.59~M_{\odot}}}%
		{GW200209}{\ensuremath{39.45~M_{\odot}}}%
		{GW200210}{\ensuremath{7.89~M_{\odot}}}%
	}[\PackageError{MDCGWMCHIRP}{Undefined option: #1}{}]
}%

\newcommand{\OTHREEGWIFOS}[1]{%
	\IfEqCase{#1}{%
		{GW200112}{L1}%
		{GW200115}{H1L1}%
		{GW200128}{--}%
		{GW200129}{H1L1V1}%
		{GW200202}{--}%
		{GW200208q}{--}%
		{GW200208am}{--}%
		{GW200209}{--}%
		{GW200210}{--}%
	}[\PackageError{OTHREEGWIFOS}{Undefined option: #1}{}]
}%

\newcommand{\OTHREEGWSNR}[1]{%
	\IfEqCase{#1}{%
		{GW200112}{\ensuremath{18.79}}%
		{GW200115}{\ensuremath{11.42}}%
		{GW200128}{--}%
		{GW200129}{\ensuremath{26.61}}%
		{GW200202}{--}%
		{GW200208q}{--}%
		{GW200208am}{--}%
		{GW200209}{--}%
		{GW200210}{--}%
	}[\PackageError{OTHREEGWSNR}{Undefined option: #1}{}]
}%

\newcommand{\OTHREEGWFAR}[1]{%
	\IfEqCase{#1}{%
		{GW200112}{\ensuremath{4.05\times10^{-4}}}%
		{GW200115}{\ensuremath{6.61\times10^{-4}}}%
		{GW200128}{\ensuremath{> \OTHREEOPA{}}}%
		{GW200129}{\ensuremath{2.11\times10^{-24}}}%
		{GW200202}{\ensuremath{> \OTHREEOPA{}}}%
		{GW200208q}{--}%
		{GW200208am}{--}%
		{GW200209}{--}%
		{GW200210}{--}%
	}[\PackageError{OTHREEGWFAR}{Undefined option: #1}{}]
}%

\newcommand{\OTHREEGWPASTRO}[1]{%
	\IfEqCase{#1}{%
		{GW200112}{\ensuremath{>0.99}}%
		{GW200115}{\ensuremath{>0.99}}%
		{GW200128}{--}%
		{GW200129}{\ensuremath{>0.99}}%
		{GW200202}{--}%
		{GW200208q}{--}%
		{GW200208am}{--}%
		{GW200209}{--}%
		{GW200210}{--}%
	}[\PackageError{OTHREEGWPASTRO}{Undefined option: #1}{}]
}%

\newcommand{\OTHREEGWMCHIRP}[1]{%
	\IfEqCase{#1}{%
		{GW200112}{\ensuremath{35.37~M_{\odot}}}%
		{GW200115}{\ensuremath{2.57~M_{\odot}}}%
		{GW200128}{--}%
		{GW200129}{\ensuremath{32.74~M_{\odot}}}%
		{GW200202}{--}%
		{GW200208q}{--}%
		{GW200208am}{--}%
		{GW200209}{--}%
		{GW200210}{--}%
	}[\PackageError{OTHREEGWMCHIRP}{Undefined option: #1}{}]
}

\newcommand{\OTHREERETRACTIONS}{23}
\newcommand{\OTHREEGSTLALRETRACTIONS}{15}

\newcommand{\RETRACTIONFAR}{\ensuremath{1.67~\mathrm{per}~\mathrm{year}}}
\newcommand{\RETRACTIONSNR}{\ensuremath{14.5}}
\newcommand{\MDCRETRACTIONFARTHRESH}{one per year}

\newcommand{\BANKMASSLOW}{\ensuremath{1.0~M_{\odot}}}
\newcommand{\BANKMASSHIGH}{\ensuremath{200~M_{\odot}}}

\newcommand{\BHMASSLOW}{\ensuremath{3.0~M_{\odot}}}
\newcommand{\NSMASSLOW}{\ensuremath{1.0~M_{\odot}}}
\newcommand{\NSMASSHIGH}{\ensuremath{3.0 M_{\odot}}}
\newcommand{\TOTALMASSHIGH}{\ensuremath{400.0~M_{\odot}}}
\newcommand{\MASSRATIOHIGH}{\ensuremath{20}}

\newcommand{\NSSPIN}{\ensuremath{0.05}}
\newcommand{\BHSPIN}{\ensuremath{0.99}}
\newcommand{\CHIPBOUND}{\ensuremath{1\times10^{-3}}}

\newcommand{\MCHIRPBOUNDARY}{\ensuremath{1.73~M_{\odot}}}
\newcommand{\LOWMCHIRPWAVEFORM}{\texttt{TaylorF2}}
\newcommand{\HIGHMCHIRPWAVEFORM}{\texttt{SEOBNRv4}}

\newcommand{\MDCELEVENLOFARLATENCY}{\ensuremath{14.58}}
\newcommand{\MDCELEVENHIFARLATENCY}{\ensuremath{10.30}}

\newcommand{\MDCTWELVELOFARLATENCY}{\ensuremath{12.04}}
\newcommand{\MDCTWELVEHIFARLATENCY}{\ensuremath{9.86}}

%% file: introduction.tex
\section{Introduction}
Ever since the first observing run (O1) of the LIGO Scientific~\cite{ligo},
Virgo~\cite{virgo}, and KAGRA~\cite{kagra} collaboration, the field of gravitational-wave (GW) astronomy has proven to be an invaluable 
tool for probing the universe. By detecting mergers of black holes and neutron stars~\cite{LIGOScientific:2018mvr, gwtc-2, gwtc-2.1, LIGOScientific:2021djp},
GW astronomy has given us the ability to study the universe in new ways.
This has led to a host of new scientific results~\cite{ligo2017gravitational, abbott2021tests, abbott2020properties}.
GW searches are the first step to producing results within GW astronomy. These results are then used by downstream tools to facilitate
multi-messenger astronomy, estimation of source parameters, and to perform studies on the population statistics and astrophysics of compact objects,
cosmology, and tests of general relativity.

Only GWs produced by the mergers of the largest compact objects in the universe like black holes and neutron stars are loud enough to be observable by 
GW detectors like the Laser Interferometer Gravitational-wave Observatory (LIGO), Virgo, and KAGRA. Even then, GW signals reaching Earth are very
faint, and heavily dominated by detector noise. Matched filtering~\cite{PhysRevD.60.022002} is the primary tool employed by modeled GW searches to detect
GW signals in noisy data. In this process, the data is cross-correlated against a template of a GW waveform predicted by general relativity, producing
a signal-to-noise ratio (SNR) timeseries as output.

GstLAL~\cite{Messick:2016aqy, Cannon:2020qnf, Sachdev:2019vvd, Hanna:2019ezx} is a stream-based GW search pipeline that has contributed to the LVK's GW detections since O1.
It implements time-domain matched filtering to recognize periods of time 
where GW signals are possibly buried in noise (called ``triggers"). It then calculates a likelihood ratio (LR)~\cite{Tsukada:2023,Cannon:2015gha, Cannon:2012zt}
as a ranking statistic for assigning significance to these triggers. The triggers with particularly high LRs are retained and called GW candidates.
Based on the LRs and rate of triggers recognized as noise, the LRs of candidates are converted to a false alarm rate (FAR), which represents our confidence in
the candidate. Other search pipelines, such as PyCBC~\cite{Dal_Canton_2021, Davies_2020, pycbc}, MBTA~\cite{mbta, Adams_2016}, SPIIR~\cite{spiir, spiir_2017},
and IAS~\cite{ias, Zackay_2021}, also use similar techniques.

The GstLAL pipeline can operate in one of two modes: a low-latency ``online" mode, or a high-latency ``offline" mode. The online mode is designed to
matched-filter the data, produce triggers, recognize candidates, and assign FARs in near-real time. The results are then immediately uploaded to
the Gravitational Wave Candidate Event Database (GraceDB)~\cite{gracedb}, from where a public alert can be sent if the upload meets certain criteria.
The ability of GstLAL to produce results in near-real time and hence serve as an independent messenger in the detection of astronomical events is
particularly useful. GW170817~\cite{GCN21505,LIGOScientific:2017vwq}, a binary neutron star merger (BNS), is an excellent example of GW searches contributing to a multi-messenger detection,
which led to many new scientific results~\cite{PhysRevLett.121.161101, PhysRevLett.123.011102}.

In contrast, the offline mode is designed to be run after all the data are available. The matched filtering, significance estimation, and FAR assignment
stages are generally done one after the other in the offline analysis, and do not need to be done simultaneously like in the online analysis.
Since the full background data can be used to assign significances to all candidates, it has traditionally been considered more sensitive than the online analysis.
Since it operates in high latency, it is resilient to any processing delays, data availability delays, and hardware downtime.
Consequently, it has also traditionally been considered more reliable and robust than the online analysis.
Because of this, the process of matched filtering, which is otherwise
identical for both operating modes, has always been repeated for the offline analysis by all search pipelines, after it was initially done in 
low-latency for the online analysis. Since GW searches are expensive, both in terms of time and computational resources, this repitition has a
significant human and computational cost. With GW searches always looking to produce more scientific results, they are becoming ever larger,
expanding to new parameter spaces, and analyzing more data than ever before. Consequently, the associated cost of running them is quickly
starting to become unfeasible.

In this work, we address whether the repitition of matched filtering, which requires the overwhelming majority of computational power and time of
any GW search, is necessary. In \secref{sec:software}, we describe the GstLAL pipeline and its two
operating modes in detail. In \secref{sec:methodology}, we introduce a novel technique in which the data products
created by the matched filtering of an online analysis are used in an offline fashion. In \secref{sec:results}, we compare the results of this technique to a traditional
offline analysis, to answer the question of whether data needs to matched-filtered a second time.

%% file: software_A.tex
\section{Software}
\label{sec:software}

\subsection{General GstLAL methods}
The GstLAL workflow, in either operating mode, contains two broad stages: a setup stage, and a data processing stage.
In the setup stage, input data products are precomputed for use during the data processing stage.
Like all modeled GW searches, GstLAL uses a ``bank" of GW waveform templates.
The template bank ahead of O4 was generated using the \texttt{manifold}~\cite{PhysRevD.108.042003} software package as described in~\cite{Sakon:2022ibh}.
It covers waveforms produced by binary mergers with component masses from 1-200$M_\odot$ and dimensionless spins up to $\pm 0.99$.
This results in a bank containing approximately 2 million total waveforms.

The template bank is split into ``template bins", each of around 1000 templates, sorted by linear combinations of their Post-Newtonian phase coefficients~\cite{Morisaki:2020oqk} as described in \cite{Sakon:2022ibh}.
The templates are additionally whitened using a power spectral density (PSD) that represents the frequency characteristics of detector noise in the data.
The templates in a single template bin are processed together for the purpose of matched filtering and background collection,
and consequently a single job within a GstLAL analysis corresponds to a specific template bin.

The next stage is the data processing stage. This involves the matched filtering process, significance assignment, FAR calculation, and uploading the results (the last only applicable for the online mode).
The SNR timeseries produced by matched-filtering the data with a particular template is defined as:

\begin{equation}
\label{eq:snr}
\mathrm{SNR}(t) = \int_{-\infty}^{\infty} d\tau \hat{d}(t + \tau) \hat{h}(\tau)
\end{equation}
where
$\hat{d}(\tau)$ is the data whitened with the PSD, and
$\hat{h}(\tau)$
is the similarly whitened template. Since matched filtering needs to be performed for the full data for every template, it is extremely computationally intensive.
Though dependent on factors like cluster availability and computational power of the processors, we can calculate a rough estimate of the percent of 
time in an offline analysis taken up by matched filtering. The duration of the first part of O4 is around 8 months. An offline analysis
over this period of time would take around 2 months. The setup stage, combined with significance assignment and FAR calculation takes around 1 day.
That means matched filtering accounts for more than 98\% of the time required for an offline analysis.

The SNR timeseries of every template is used to form triggers, by identifying times when the SNR exceeds the threshold value of 4.
Parallelly, background data is also collected by identifying triggers that originate from noise~\cite{Joshi:2023ltf}.
The background data thus collected is then used to rank the triggers using the likelihood ratio as described in~\cite{Tsukada:2023}.
Triggers with a high LR are retained as GW candidates, and the LRs of noise triggers, as well as the livetime of the analysis is used to
convert the LRs of candidates into FARs.

In the following subsections, we will discuss how the implementation of these steps differs for the online and offline
operating modes of the GstLAL analysis.

%% file: software_B.tex
\subsection{Online GstLAL Analysis}

The online analysis is designed to ingest the data coming from the detectors in real-time, and produce results
with minimal delay.
While running an online analysis, if the FAR of a candidate crosses a certain threshold~\cite{opa}, it gets uploaded to GraceDB.
Typically, this happens within 10-20 seconds of the GW reaching Earth~\cite{Ewing:2023}.
If certain other criteria are met, a skymap is generated showing the likely sky location of the source of the
GW~\cite{Singer:2015ema, Singer:2016eax, Ashton:2018jfp, Romero-Shaw:2020owr}, and a public alert is issued~\cite{gcn}.
Additional information, such as low-latency parameter estimation of the source of the candidate~\cite{rose2024rapid}, and the probability
of astrophysical origin for different source classes~\cite{Ray:2023nhx, villa2022astrophysical, Andres_2022} is also included in the public alert.
Astronomers
can then choose to follow up on this alert, and correlations can be made with other messengers~\cite{piotrzkowski2022searching, urban2016monsters, cho2019low, 2016ApJ...826L..13A}.
In this way, the online analysis plays an instrumental role in multi-messenger astronomy.

To facilitate this, the online analysis needs to ensure the following two principles are observed:
\begin{enumerate}
\item{Causality: To process a GW candidate, the analysis can only use data available up to that point in time. It cannot wait for more data to become available in the future.}
\item{Keeping up with live data: The analysis cannot fall behind the incoming data. If it does, it needs to drop some data to catch up.}
\end{enumerate}

Currently, the GstLAL analysis has the ability to measure the PSD of the data in small batches of 4 seconds, and whiten the data accordingly,
thus effectively ensuring causality well enough to serve the near-real time search.
However, this ability does not extend to template whitening, and the templates need to be whitened during the setup stage.
As a result, the online analysis uses a PSD projected to represent future noise, to whiten the templates.
To minimize this effect, every week in O4, the GstLAL team has been whitening the templates used by the online analysis using the PSD measured over the previous
week's data. The expectation is that this captures any changes in the noise characteristics with a timescale equal to or larger than a week.

Additionally, to ensure causality, in order to rank a particular trigger, the online analysis can only use background data collected up to the point
of that trigger. It cannot wait for the full background data to be collected. As a result, it may happen that the background used to rank a specific
trigger has not converged fully, and so the LR and FAR assigned to that trigger are not as accurate as they would have been had we waited for the 
full background to be accumulated. Hence, the online analysis is traditionally considered to be less reliable.

Finally, to ensure the principle of keeping up with live data, if for any reason the analysis is unable to process some stretch of data, that data is permanetly lost.
The analysis cannot go back and re-analyze the data. Data drops like this generally happen for one of three reasons:

\begin{enumerate}

\item{If the analysis hits a period of high latency, either due to its own internal processing, or due to external reasons like live data delivery failures, the analysis
drops this data and moves on to newer data after waiting for a set amount of time (60 seconds).}
\item{During the running of the analysis, there will be regularly scheduled maintenance, both internally for the analysis, and externally for the hardware it is running on.
It is attempted to make these periods of downtime as short as possible, and to make them coincide with periods of time when the detectors are not producing data.}
\item{Unintentional hardware downtime}
\end{enumerate}
Due to these reasons, the online analysis is traditionally considered to be less sensitive and robust as compared to an offline analysis.

%% file: software_C.tex
\subsection{Offline GstLAL Analysis}
The offline analysis is designed to be robust, reliable, and more sensitive than the online analysis.
It is typically used to generate more in-depth results than the online analysis, such
as studies on population properties of compact binary systems~\cite{Abbott_2021,KAGRA:2021duu}, and tests
of general relativity~\cite{PhysRevD.100.104036, PhysRevD.103.122002, PhysRevLett.116.221101, PhysRevLett.123.011102}.
To do this, offline analyses will commonly be run with simulated gravitational wave signals injected into the data. These
are called ``injections", and they are used to measure the response of the analysis (i.e. the sensitivity)
for GWs from sources in various parameter
spaces, at different distances, etc.

The analysis is performed after the full data become available, and so there are no latency
constraints. As a result, the analysis does not need to adhere to the principles of causality
and keeping up with live data, like the online analysis did.

Consequently, the PSD used for template whitening can be directly measured
from the full data itself, guaranteeing the best representaion of detector noise. In practice, the data are
divided into week-long chunks, and the process of PSD measurement and template whitening is done separately
for every chunk. This further improves how well we capture detector noise, and increases sensitivity.

Similarly, the filtering and ranking stages (which involves significance and FAR assignment), do not need to occur simultaneously for a given
trigger in the offline analysis, and each stage can be done for all triggers before moving on to the next.
This means that during the ranking stage, the background data from the full analysis
can be used, leading to more reliable results, as well as an increase in sensitivity. With this in mind,
the matched filtering stage and the ranking stage of the 
GstLAL offline workflow are completely modular, and designed to be run independently.

Additionally, the analysis does not need to drop data if it
gets affected by some source of latency, whether it is a latency in its internal data processing or
a disruption in data delivery. This means that the analysis is guaranteed to 
process all available data without dropping any like the online analysis, making it more reliable and sensitive.
Furthermore, the analysis
can make use of more robust data quality and data veto information, provided by external high-latency tools. This is thought to make the analysis
more sensitive by removing obviously bad data that would have otherwise created false positives in both
the candidates and the background.

%% file: methodology.tex
\section{Methodology}
\label{sec:methodology}

In the descriptions of the online and offline GstLAL analyses above, the only differences between the
matched filtering stages of the two are PSDs used for template whitening,
and the fact that the two might not analyze exactly the same set of data.
Since we expect the projected PSD used by the online analysis to still be a good approximation of the true
PSD measured over the data that the offline analysis uses, the effect of PSD mismatch should be low~\cite{Messick:2016aqy}.
Additionally, the weekly whitening of the online analysis templates using the previous week's PSD will
lower any SNR loss due to PSD mismatch further. Similarly, with improvements to the stability of the
online analysis~\cite{Ewing:2023}, data distribution, and computing hardware done before O4, we expect the
effect of data drops to also be low. In \secref{sec:results} we show that the online analysis only drops
around 5\% data as compared to the offline analysis, and also discuss ways to make up this lost 5\%.

\subsection{Online Rank}
Based on this, we developed a novel technique that takes the data products created by the matched filter
stage of an online analysis (i.e. triggers and background data), and replaces the offline analysis'
matched filter stage with these. The rest of the offline analysis (i.e. the ranking stage),
is kept the same. This is possible since the two stages are designed to be modular, as described in \secref{sec:software}.
We call this technique an ``online rank", since the matched
filtering is taken from the online analysis, and an offline ranking stage is added to it.

Alongside the modularity of the GstLAL offline workflow, the key to making an online rank possible is a feature of the GstLAL online analysis, called ``snapshotting".
Every 4 hours, each job in the the online analysis (each corresponding to a template bin) will write a 
snapshot of the triggers and background data it has collected to disk. The trigger snapshot files are discrete, i.e.
each file will only contain triggers created since the previous snapshot.
In contrast, the background snapshot files are cumulative, i.e. each file will contain background data since
the start of the analysis to the time of the snapshot. Snapshotting can also be used to save the progress of the
online analysis in case something goes wrong and the analysis needs to be restored to a working state.

\begin{figure*}
\includegraphics[width=\textwidth]{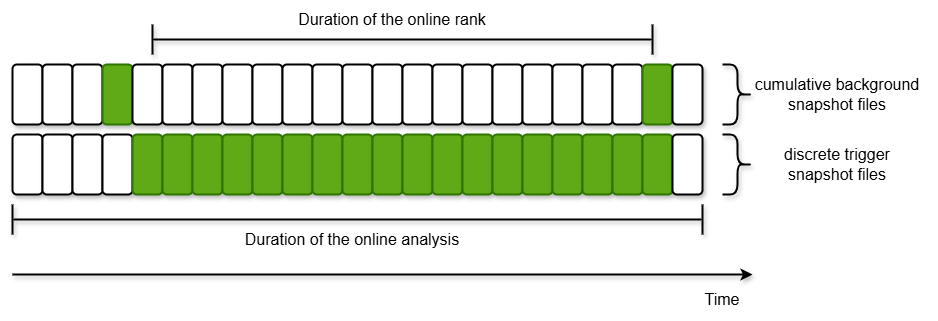}
\caption{\label{fig:schematic}
A schematic showing which files are selected for an online rank, from the online analysis.
The rectangles represent the two types of data products created by the online matched
filtering process. The rectangles at the bottom represent the trigger snapshot files, and those
at the top represent the background snapshot files. Each snapshot file is 4 hours long. The
rectangles colored green are the ones selected by the online rank.
Since the trigger snapshot files are discrete, all those having an overlap with the online
rank duration are selected. In contrast, since the background snapshot files are cumulative,
the earliest one containing all the background data for the duration of the online rank, and the
latest one containing of it are chosen. These two are then subtracted to produce a background
file containing exactly the background data for the duration of the online rank, to the
granularity of the 4 hour snapshots. This process is repeated for every template bin in the analysis.
}
\end{figure*}

The setup process for an online rank involves going through all the trigger and background snapshot files
written by the online analysis, and picking the relevant files to forward to the rank stage. The
user can specify a start and end time for the online rank, which defines the duration of the online rank.
Since trigger snapshot files are discrete, every trigger file whose duration (which is encoded in the filename)
has an overlap with the duration of the online rank is forwarded to the rank stage. For the background 
snapshot files, however, since they are cumulative, the earliest snapshot file that contains all the background
data of the duration of the online rank is chosen.
If the start time of the online analysis is different from the start time of the online rank, the latest background
snapshot file that doesn't overlap at all with the duration of the online rank is also chosen. This is subtracted from
the earlier file, to produce a background file that exactly contains the background data for the duration of the
online rank, to the granularity of the 4 hour snapshots.
This procedure is illustrated in \figref{fig:schematic}.
Typical of an offline analysis, this background file containing the full background
data for the duration of the online rank is used to rank every trigger, leading to more reliable and sensitive results.
Since trigger and background files are processed
separately for every template bin, this process is repeated for every template bin. In this way, relevant files
can be extracted from an online analysis that will typically be running for the full observing run, and offline results
for a subset of the duration can be calculated from them. Since the snapshotting
interval of 4 hours is relatively small as compared to typical analysis periods of many months, trigger
and background data can be extracted from the online analysis for offline processing with high precision.

In order to make the online rank results even more reliable and sensitive,
we can augment the inputs to the online rank with triggers and background data
from times that the online analysis dropped. Specifically, for every job we
calculate such a ``dropped data segments", which an offline filtering would have
analyzed but the online filtering did not, and set up a traditional offline
analysis using these dropped data segments.  By combining the online rank inputs
with the results of the dropped data refiltering analysis, we can be sure the
online rank produces offline results for exactly the same periods of time that
the traditional offline analysis would have. The typical amount of dropped data
for any job is around 5\% of the total time covered by the offline
segments. This is demonstrated in \secref{sec:results}. Additional details can
also be found in~\cite{gwtc4_gstlal_method}.

\subsection{Offline Rank Stage}
After relevant trigger and background files for every template bin are chosen from the online analysis, they
are forwarded to the offline rank stage. From this point on, the online rank analysis proceeds identically
to the traditional offline analysis. Details of how the offline rank stage works can be found in~\cite{gwtc4_gstlal_method}.

\subsection{Computational Cost Reduction}
The online rank procedure requires an online analysis to have been run over the relevant period of data first.
It enables us to re-use the matched filtering data products from the online analysis in order to get
offline results. Since matched filtering is the bulk of the computational cost of any modeled GW search (98\%,
as discussed in \secref{sec:software}), over the course of an observing run (i.e. including running analyses
to get both online and offline results), the online rank procedure represents approximately a 50\% reduction
in the total computational cose.

Given that the online analysis has alrady been run, we can calculate the time saved to get offline results via an online rank. Since
the time required for matched filtering scales linearly with the amount of data analyzed, but the time required
for an online rank does not strongly depend on the amount of data analyzed, the time saved because of the online
rank method depends on the length of the analysis. The duration of the first part of O4 is around 8 months. The
time required to perform a traditional offline analysis with injections over this period of time
is approximately 4 months. We can get offline results for the same duration of time via an online rank
in as low as 5 hours, if dropped data refiltering is not included.
This represents a 99.8\% reduction in the computational time in the best case scenario.

As discussed before, the typical amount of data dropped by an online analysis is 5\%.
This means that even if we choose to perform the 
dropped data refiltering analysis to augment the online rank, we still get approximately a 95\% reduction
in the amount of time required to get offline results.

%% file: results_A.tex
\section{Results}
\label{sec:results}
\subsection{MDC Data Set and Analyses}
In order to test the efficacy of our new method, we ran an online analysis over a mock data challenge (MDC).
This involved running the analysis over 38 days of O3 data from the LIGO Hanford and Livingston detectors,
as well as the Virgo detector. The data extended from 7 January 15:59:42 UTC 2020 to 14 February 20:39:42 UTC 2020.
These data were then shifted in time by 125952000 seconds to extend from 4 January 10:39:42 UTC 2024 to 11 February
15:19:42 UTC 2024, to make them appear as though they were live data when we ran the online analysis.
The MDC also involved an injection campaign. More details about the MDC, including
details about the injection set used can be found in~\cite{mdc_analytics}.

We then performed an online rank on this analyis, as well as a traditional offline analysis on the same amount of
data. After accounting for the times when no detectors were producing data, we find that the online rank had
a livetime of 34.41 days, whereas the offline analysis had a livetime of 36.05 days. This means that over the 
course of 38 days, the online analysis dropped around 4.5\% of the data. To compensate for this, we also performed
a dropped data refiltering analysis over the 4.5\% dropped data. All of these analyses were performed using 
the GstLAL O4 template bank, described in \secref{sec:software}.

%% file: results_B.tex
\subsection{Sensitivity Comparisons}
\begin{figure}
\includegraphics[width=\columnwidth]{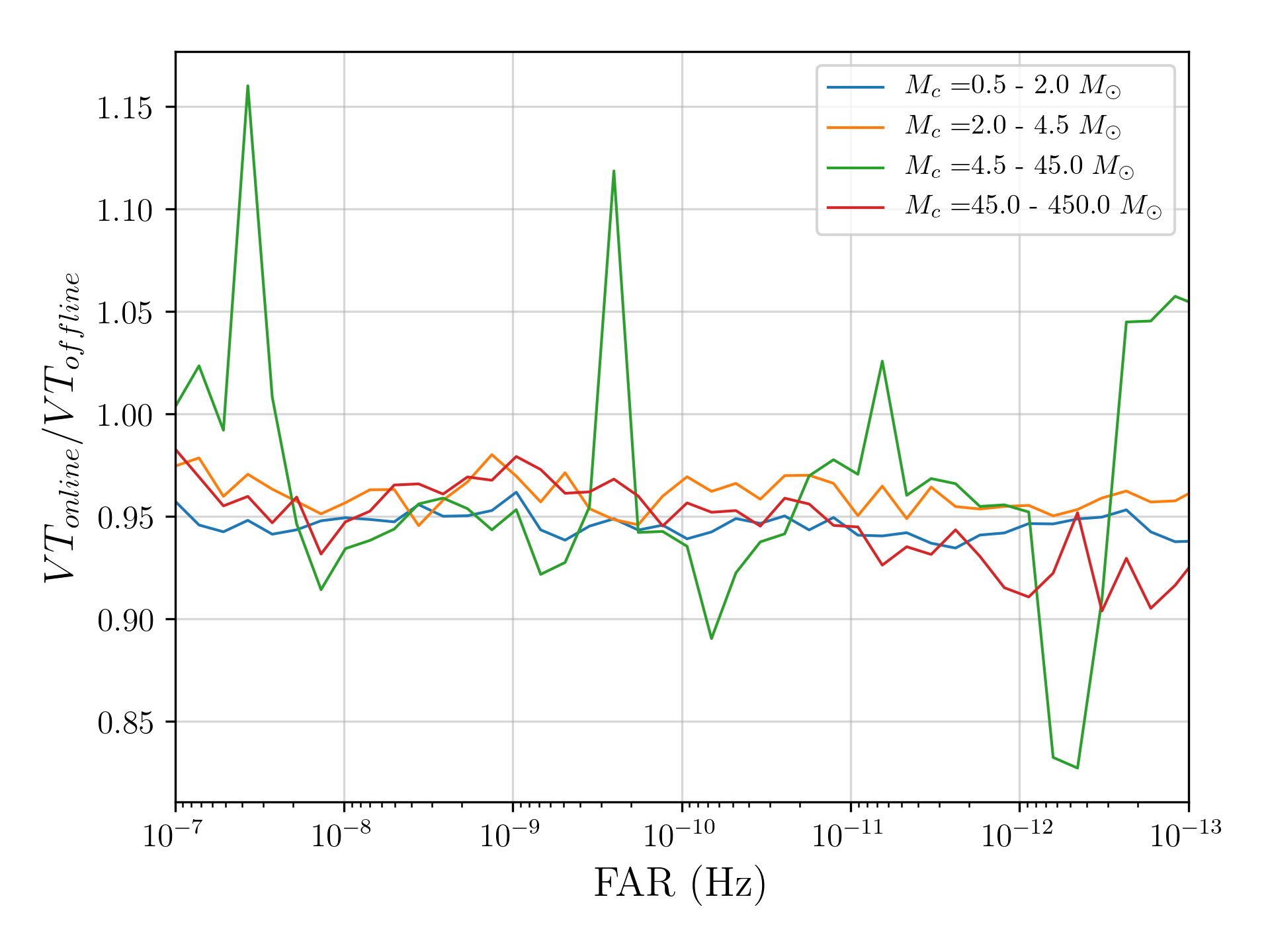}
\caption{\label{fig:vt_ratio_raw}
The ratio of the sensitive volume-times of the online rank to that
of a traditional offline analysis over the same period of time, calculated for different mass bins and at different FAR thresholds. The
fact that the \textit{VT} ratios for all mass bins are close to 1 across FAR thresholds
tells us that the online rank method is very close in sensitivity
to a traditional offline analysis. The 5\% loss in VT comes from
the fact that the online analysis dropped approximately that much data.
The peaks and troughs in the BBH line are because of the small
number of statistics in that mass bin.
}
\end{figure}

To compare the sensitivities of the online rank and the offline analysis, we can calculate
the \textit{VT}s of both, and then take the ratio of the two. Here, we have calculated the 
\textit{VT} separately for injections with chirp mass in four different mass bins, roughly
corresponding to four source categories: binary neutron star mergers (BNS, chirp mass between 0.5 to 2 $M_\odot$)
neutron star-black hole mergers (NSBH, chirp mass between 2 to 4.5 $M_\odot$),
binary black hole mergers (BBH, chirp mass between 4.5 to 45 $M_\odot$), and 
intermediate-mass black hole mergers (IMBH, chirp mass between 45 to 450 $M_\odot$).
The \textit{VT} is calculated at different FAR thresholds for considering an injection
to be found by the analysis.
The results of this \textit{VT} comparison for different mass bins and FAR thresholds, for the pure online rank and offline analysis
can be seen in \figref{fig:vt_ratio_raw}. It shows us that the online rank is almost as
sensitive as the offline analysis. We note that the 5\% loss in the online rank \textit{VT}
as compared to the offline analysis lines up perfectly
with the 5\% of data dropped by the online analysis.

\begin{figure}
\includegraphics[width=\columnwidth]{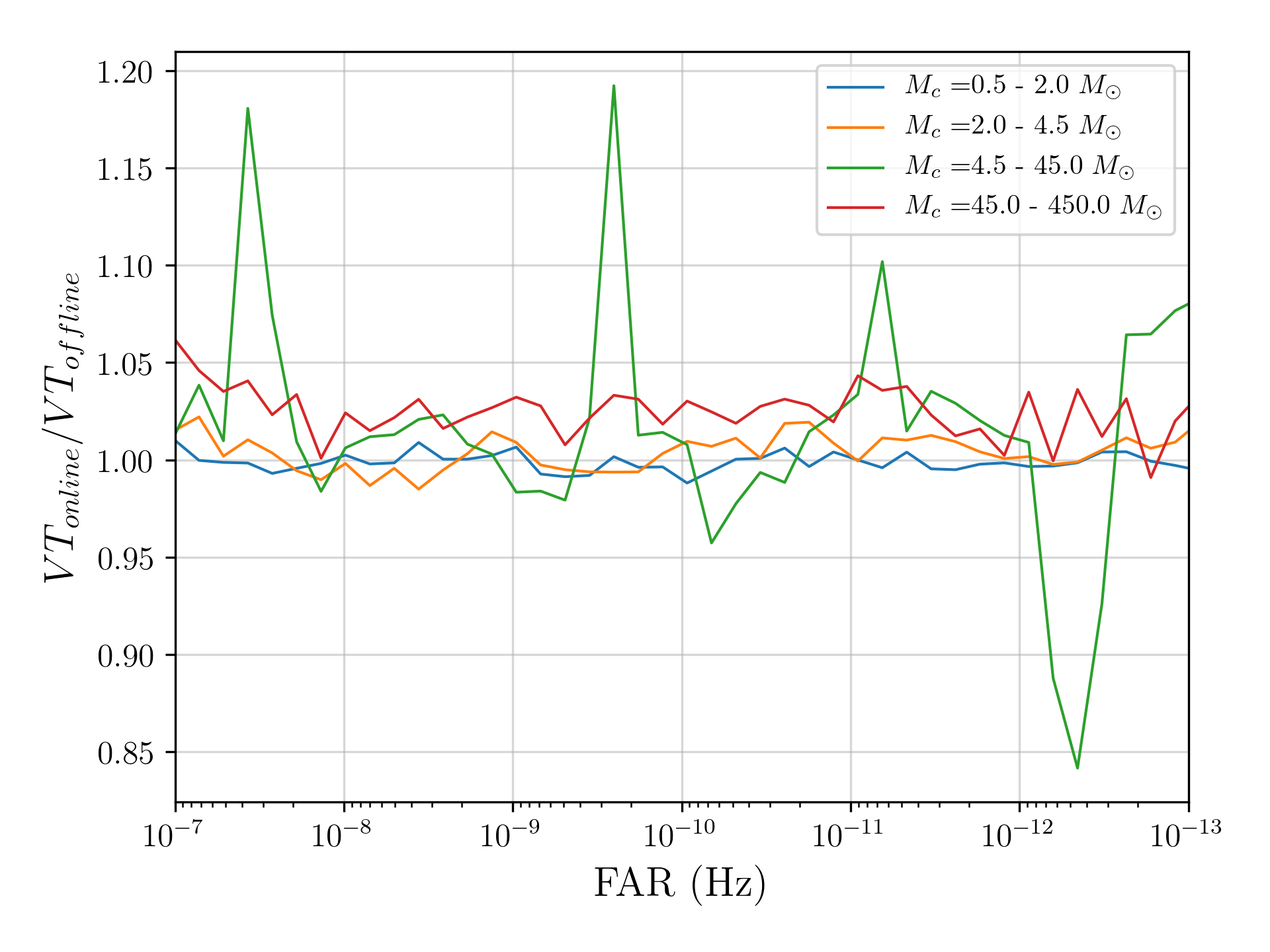}
\caption{\label{fig:vt_ratio_dropped_data_refiltering}
The ratio of the sensitive volume-times of the online rank augmented with
triggers and background data from the periods of time dropped by the online analysis to that
of a traditional offline analysis over the same period of time. We
see that the 5\% loss in \textit{VT} seen in \figref{fig:vt_ratio_raw}
is recovered by adding the 5\% of dropped data. This shows that
the online rank method is exactly as sensitive as a traditional offline analysis.
}
\end{figure}

Next, we repeat the procedure for the online rank augmented with
the dropped data refiltering analysis, and compare its \textit{VT}
to the \textit{VT} of the offline analysis. The result of this is shown
in \figref{fig:vt_ratio_dropped_data_refiltering}. The fact that the 
\textit{VT} ratios are now much closer to 1 tells us that the previous 5\% loss
in \textit{VT} was indeed coming from dropped data, and that by augmenting
an online rank with a dropped data refiltering analysis, we can get
offline results that are as sensitive as a traditional offline analysis
in a fraction of the time.

%% file: results_C.tex
\subsection{Candidate Lists}
Next, we compare the candidate lists from the online rank augmented with dropped data refiltering and offline analysis
as a further check on the reliability and sensitivity of the online rank. There are 
9 previously reported GW signals in the MDC data, and both analyses are able to
detect 5 of them with a FAR of 1/month ($3.86  \times 10^{-7}$ Hz) or less.
The candidate list for the online
rank is shown in \tabref{tab:online}, and that for the offline analysis is shown
in \tabref{tab:offline}. The top 10 candidates from both analyses are the same.
Additionally, both analyses recover those 10 candidates with exactly the same template,
as evidenced by the fact that they have the same primary and secondary masses ($m_1$ and $m_2$),
as well as the same dimensionless spins ($a_1$ and $a_2$).
Since the analyses were performed on O3 data shifted in time by 125952000 seconds,
the reported times of the candidates do not match the times reported in
the third Gravitational-Wave Transient Catalog~\cite{LIGOScientific:2021djp}

\begin{table*}[]
    \centering
    \begin{tabular}{p{0.4in}p{1in}p{1.25in}p{0.5in}p{0.5in}p{0.5in}p{0.5in}}
        \textbf{Rank} & \textbf{FAR} (Hz) & \textbf{Time (UTC)} & $\mathbf{m_{1}}$ ($M_{\odot}$) & $\mathbf{m_{2}}$ ($M_{\odot}$) & $\mathbf{a_{1}}$ & $\mathbf{a_{2}}$ \\
        \hline
        1 & $5.45 \times 10^{-34}$ & 2024-01-26 01:34:58 & 40.86 & 30.5 & 0.05 & 0.05  \\
        2 & $1.43 \times 10^{-13}$ & 2024-01-11 23:03:09 & 5.24 & 1.77 & -0.29 & -0.29 \\
	3 & $6.58 \times 10^{-13}$ & 2024-01-24 21:00:11 & 59.52 & 57.08 & 0.17 & 0.17  \\
	4 & $1.15 \times 10^{-12}$ & 2024-02-05 07:41:17 & 50.36 & 34.57 & -0.2 & -0.2 \\
	5 & $8.69 \times 10^{-9}$ & 2024-02-06 03:34:52 & 50.36 & 40.86 & -0.08 & -0.08 \\
	6 & $1.20 \times 10^{-7}$ & 2024-02-04 08:49:54 & 176.4 & 184.0 & 0.6 & 0.6 \\
	7 & $4.16 \times 10^{-7}$ & 2024-01-26 06:22:45 & 70.35 & 79.75 & 0.45 & 0.45 \\
	8 & $9.95 \times 10^{-7}$ & 2024-01-17 21:57:48 & 79.75 & 59.52 & -0.02 & -0.02 \\
	9 & $1.2 \times 10^{-6}$ & 2024-02-06 03:40:07 & 40.86 & 42.6 & 0.73 & 0.73 \\
	10 & $1.46 \times 10^{-6}$ & 2024-01-29 08:21:54 & 126.3 & 57.08 & -0.08 & -0.08 \\
    \end{tabular}
    \caption{The candidate list of the online rank. The first five candidates correspond to the
previously reported events of GW200129\_065458, GW200115\_042309, GW200128\_022011, GW200208\_130117, and GW200209\_085452.
However, the times are different than those reported in~\cite{LIGOScientific:2021djp}, because the data was shifted in time.
The candidates and parameters reported by the online rank are identical to those reported by the 
traditional offline analysis in \tabref{tab:offline}}
    \label{tab:online}
\end{table*}

\begin{table*}[]
    \centering
    \begin{tabular}{p{0.4in}p{1in}p{1.25in}p{0.5in}p{0.5in}p{0.5in}p{0.5in}}
	\textbf{Rank} & \textbf{FAR} (Hz) & \textbf{Time (UTC)} & $\mathbf{m_{1}}$ ($M_{\odot}$) & $\mathbf{m_{2}}$ ($M_{\odot}$) & $\mathbf{a_{1}}$ & $\mathbf{a_{2}}$ \\
        \hline
        1 & $7.92 \times 10^{-32}$ & 2024-01-26 01:34:58 & 40.86 & 30.5 & 0.05 & 0.05 \\
        2 & $1.2 \times 10^{-13}$ & 2024-01-11 23:03:09 & 5.24 & 1.77 & -0.29 & -0.29 \\
        3 & $4.49 \times 10^{-13}$ & 2024-01-24 21:00:11 & 59.52 & 57.08 & 0.17 & 0.17  \\
        4 & $1.00 \times 10^{-12}$ & 2024-02-05 07:41:17 & 50.36 & 34.57 & -0.2 & -0.2 \\
        5 & $4.56 \times 10^{-9}$ & 2024-02-06 03:34:52 & 50.36 & 40.86 & -0.08 & -0.08 \\
        6 & $2.00 \times 10^{-7}$ & 2024-02-04 08:49:54 & 176.4 & 184.0 & 0.6 & 0.6  \\
        7 & $4.92 \times 10^{-7}$ & 2024-01-26 06:22:45 & 70.35 & 79.75 & 0.45 & 0.45  \\
        8 & $1.18 \times 10^{-6}$ & 2024-01-17 21:57:48 & 79.75 & 59.52 & -0.02 & -0.02 \\
        9 & $1.33 \times 10^{-6}$ & 2024-02-06 03:40:07 & 40.86 & 42.6 & 0.73 & 0.73 \\
        10 & $1.88 \times 10^{-6}$ & 2024-01-29 08:21:54 & 126.3 & 57.08 & -0.08 & -0.08 \\
    \end{tabular}
    \caption{The candidate list of the offline analysis. The first five candidates correspond to the
previously reported events of GW200129\_065458, GW200115\_042309, GW200128\_022011, GW200208\_130117, and GW200209\_085452.
However, the times are different than those reported in~\cite{LIGOScientific:2021djp}, because the data was shifted in time.
The candidates and parameters reported by the online rank in \tabref{tab:online} are identical to those reported by the
traditional offline analysis here}
    \label{tab:offline}
\end{table*}

Out of the 9 previously reported GW candidates in this data, the ones not found significantly by either analyses are GW200208\_222617,
GW200112\_155838, GW200202\_154313, and GW200210\_092254. The first of these was not found significantly by GstLAL
even in~\cite{LIGOScientific:2021djp}, and as such we do not expect it to be found significantly in either of the analyses performed
here. The remaining three were recovered either as single detector candidates, or recovered confidently in only one detector. Because
of the way GstLAL collects background data~\cite{Joshi:2023ltf}, these candidates were added to the background, and hence were
downweighted, resulting in them not being recovered significantly. The method described in~\cite{Joshi:2023ltf} is designed to prevent
exactly this situation. However, it is currently only compatible with the online analysis (and hence the online rank), and in order
to do a fair comparison, this method was not used for the analyses performed here. However, we did verify that by using this method
in an online analysis, the online rank is able to recover all 3 of these candidates significantly. Since this method is used for O4,
we expect GstLAL's online ranks for O4 to be immune against such scenarios.

%% file: results_D.tex
\subsection{Injection Parameter Recovery Comparisons}
\begin{figure}
\includegraphics[width=\columnwidth]{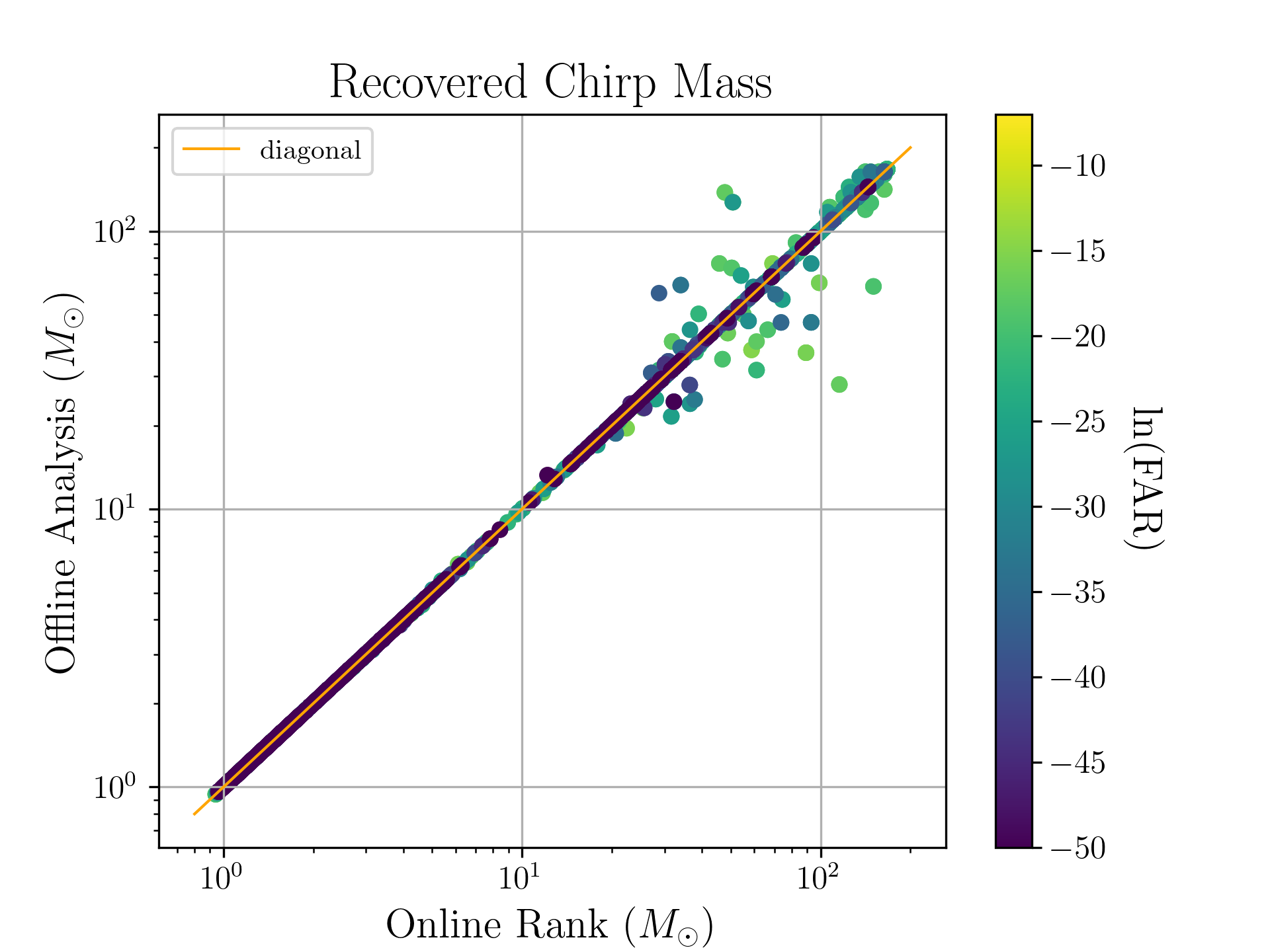}
\caption{\label{fig:mc} Scatter plot of chirp mass values recovered by the offline analysis (y-axis) and online rank (x-axis). Out of 6910 injections passing the FAR threshold of 1 per month, the majority of points land on the digonal. This indicates a consistent chirp mass recovery between the offline analysis and online rank. Deviations are seen for points with higher FARs and higher chirp masses.}
\end{figure}

\begin{figure}
\includegraphics[width=\columnwidth]{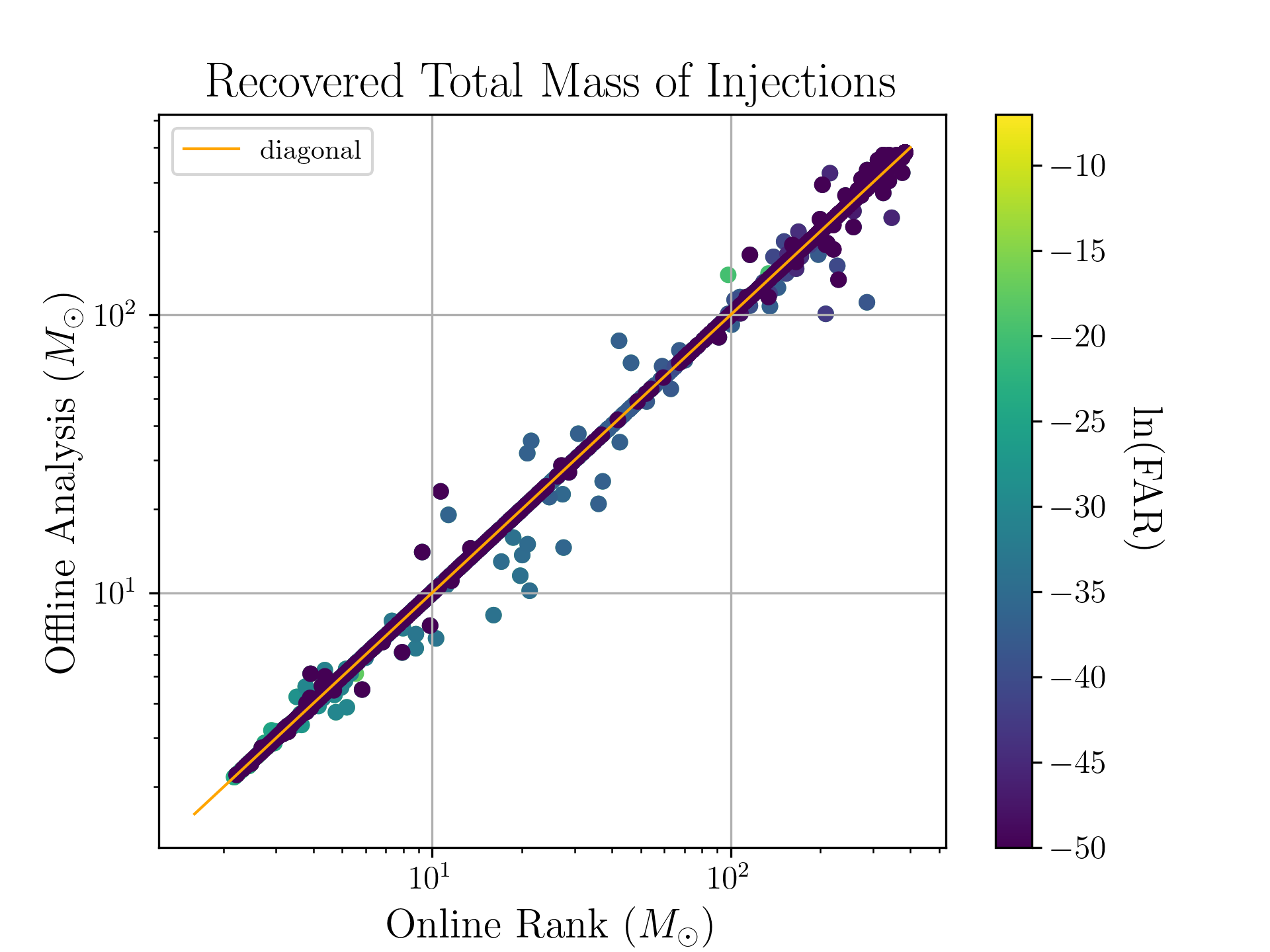}
\caption{\label{fig:mtot} Scatter plot of total mass values recovered by the offline analysis (y-axis) and online rank (x-axis). Out of 6910 injections passing the FAR threshold of 1 per month, only a few points land off the diagonal, indicating a consistent total mass recovery between the offline analysis and online rank. Total mass recovery of modelled GW searches is known to be worse than chirp mass recovery, leading to the higher scatter of points in this plot as compared to \figref{fig:mc}}
\end{figure}

\begin{figure}
\includegraphics[width=\columnwidth]{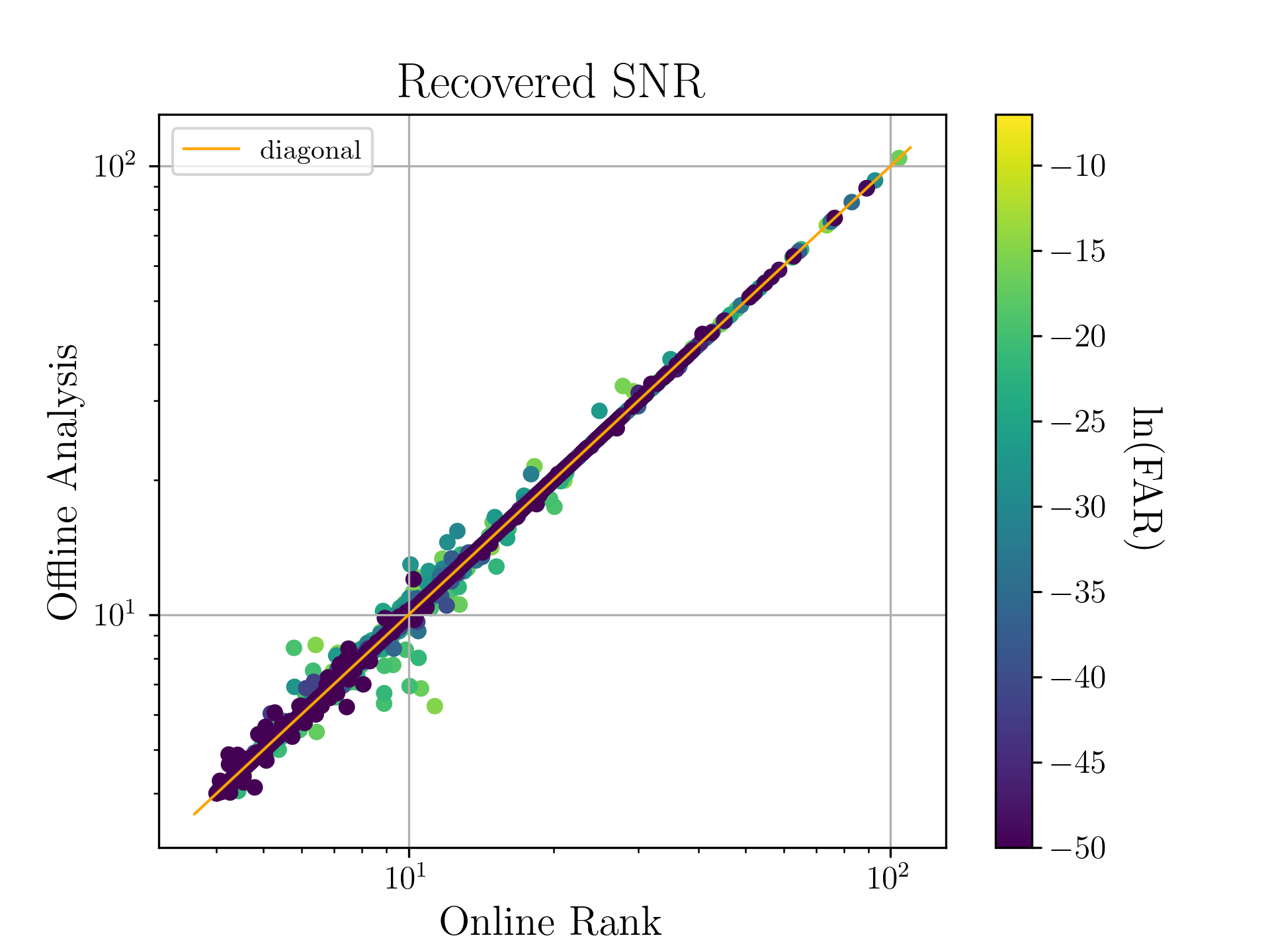}
\caption{\label{fig:snr} Scatter plot of SNR recovered by the offline analysis (y-axis) and online rank (x-axis). Out of 6910 injections passing the FAR threshold of 1 per month, most land very close to the diagonal, with a larger scatter seen at lower SNRs, and higher FARs. The small deviations from the diagonal are caused by randomness introduced by differences in the PSD used at any given time by the online and offline analyses. The fact that the points symmetrically scatter around the diagonal indicates a consistent SNR recovery between offline and online rank methods.}
\end{figure}

\begin{figure}
\includegraphics[width=\columnwidth]{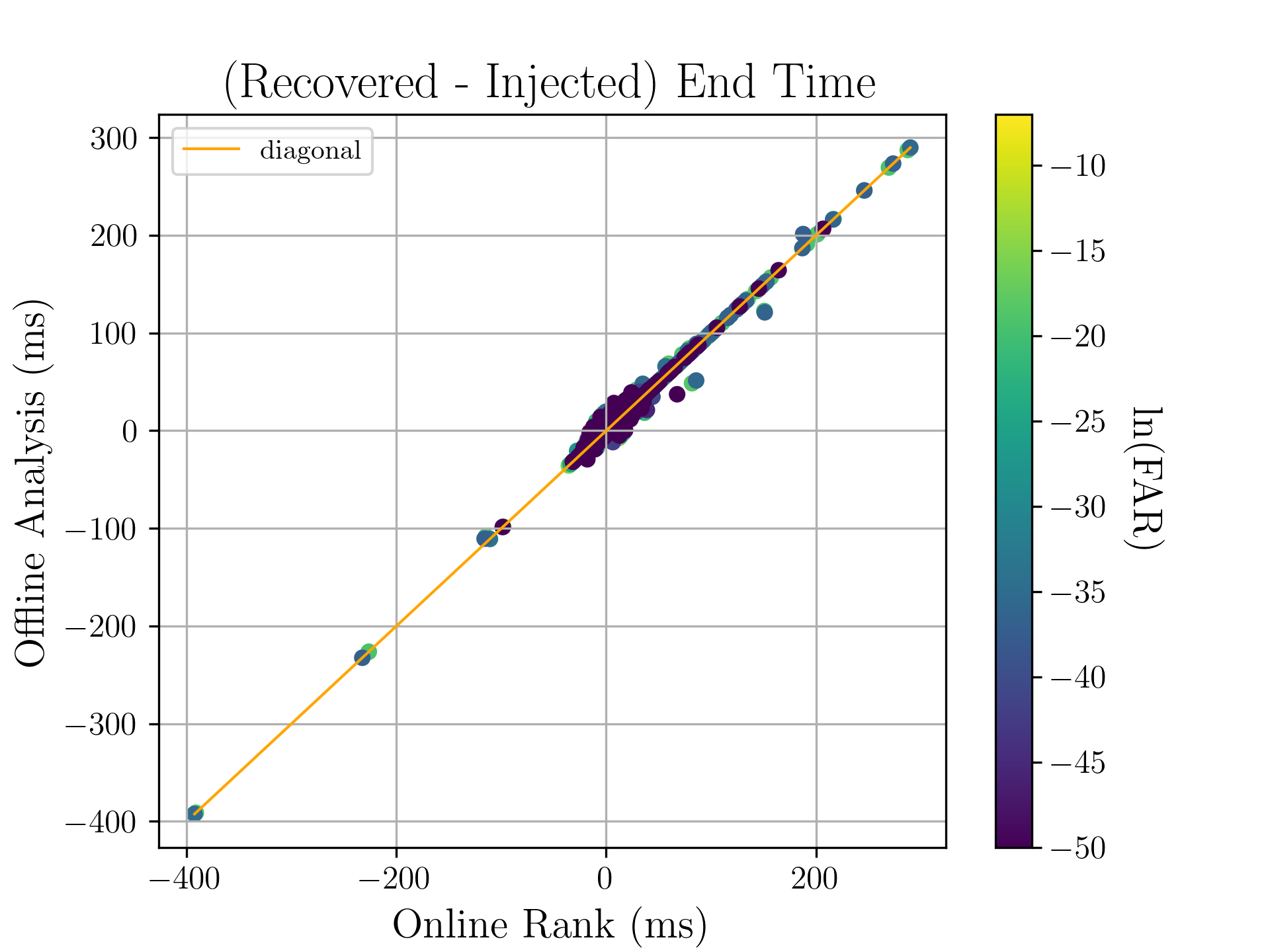}
\caption{\label{fig:time} Scatter plot of error in recovered coalescence time values by the offline analysis (y-axis) and online rank (x-axis). Out of 6910 injections passing the FAR threshold of 1 per month, almost all of the points land on the digonal, indicating a consistent end time recovery between offline and online rank methods. A few points have a high error in both analyses, reflecting bad injections recovery due to transient noise sources.}
\end{figure}

Since an injection campaign was conducted for both the online analysis (and hence the online rank),
and the offline analysis, we can compare the parameters with which the online rank and offline analysis
recovered injections. The FAR threshold used here for an injection to qualify as ``found" is 1/month
($3.86  \times 10^{-7}$ Hz). The parameters we compare here are the chirp mass, total mass, SNR, and 
coalescence time. Scatter plots of the values of these parameters recovered by the online rank and the offline analysis are
shown in \figref{fig:mc}, \figref{fig:mtot}, \figref{fig:snr}, and \figref{fig:time} respectively.
In each figure, we see that almost all points are on the diagonal, with no systematic errors.
This is further evidence that an online rank is very similar to a traditional
offline analysis.

%% file: conclusion.tex
\section{Conclusion}

In this work, we have decribed how a GstLAL analysis functions, and discussed
the differences between the online and offline modes of operation.
We introduced a new method called an online rank, in which the data products created by the matched filtering
stage of an online analysis (i.e. triggers and background data), which are saved as 
4 hour snapshots of the online analysis, can be taken and processed in an offline
fashion. This removes the drawback that the online analysis has, of not having the 
full background information available while ranking triggers. Since matched filtering
takes up the large majority of time required for an offline analysis, by not repeating
the process of matched filtering, and taking the matched filtering results from the
online analysis instead, we can get reliable and sensitive offline results
in a fraction of the time compared to what is required for a traditional offline analysis.
Over the course of an observing run (i.e. including both online and offline results),
this represents a 50\% reduction in total computational cost.

Furthermore, we discussed a technique called dropped data refiltering in which
the matched filter outputs of the online analysis are augmented by a small
offline analysis which analyzes times dropped by the online analysis. This ensures
an online rank analyzes exactly the same period of time as an offline analysis.

To test our method, we performed an online analysis on 38 days of LIGO and VIRGO
O3 data. We found that the online analysis had dropped around 5\% data as compared
to the offline analysis, consequently suffering a 5\% loss in \textit{VT}.
By adding the dropped data refiltering outputs to the online rank, we were able
to show the online rank is exactly as sensitive and reliable as a traditional
offline analysis.

Due to the significant reductions in computational effort and time enabled by online rank method, we
believe the future of GW searches lies in this paradigm, where in order to make detections in near-real
time as well as produce more detailed results for the catalog, the data is matched filtered only once per observing run, by the online analysis.
For O4, the GstLAL group has already adopted the online rank method, enabling fast
and reliable offline results, as well as fast testing on new development work.

%% file: ack.tex
This research has made use of data, software and/or web tools obtained from the
Gravitational Wave Open Science Center (https://www.gw-openscience.org/ ), a
service of \ac{LIGO} Laboratory, the \ac{LSC} and the Virgo
Collaboration.  
We especially made heavy use of the \ac{LVK} Algorithm
Library. 
\ac{LIGO} was constructed by the California Institute of Technology and the 
Massachusetts Institute of Technology with funding from the United States 
National Science Foundation (NSF) and operates under cooperative agreements 
PHYS-$0757058$ and PHY-$0823459$.
In addition, the Science and Technology Facilities Council (STFC) of the United 
Kingdom, the Max-Planck-Society (MPS), and the State of Niedersachsen/Germany 
supported the construction of \ac{aLIGO} and construction and operation of the 
GEO600 detector. 
Additional support for \ac{aLIGO} was provided by the Australian Research Council.  
Virgo is funded, through the European Gravitational Observatory (EGO), by the 
French Centre National de Recherche Scientifique (CNRS), the Italian Istituto 
Nazionale di Fisica Nucleare (INFN) and the Dutch Nikhef, with contributions by 
institutions from Belgium, Germany, Greece, Hungary, Ireland, Japan, Monaco, 
Poland, Portugal, Spain.

This material is based upon work supported by NSF's LIGO Laboratory which is a major facility fully funded by the National Science Foundation.
The authors are grateful for computational resources provided by
the \ac{LIGO} Lab culster at the \ac{LIGO} Laboratory and supported by 
PHY-$0757058$ and PHY$-0823459$, the Pennsylvania State University's Institute 
for Computational and Data Sciences gravitational-wave cluster, 
and supported by 
OAC-$2103662$, PHY-$2308881$, PHY-$2011865$, OAC-$2201445$, OAC-$2018299$, 
and PHY-$2207728$.  
LT acknowledges support from the Nevada Center for Astrophysics.
CH Acknowledges generous support from the Eberly College of Science, the 
Department of Physics, the Institute for Gravitation and the Cosmos, the 
Institute for Computational and Data Sciences, and the Freed Early Career Professorship.
MWC acknowledges support from the NSF with grant numbers PHY-2308862 and PHY-2117997.
US and SS acknowledge support from NSF PHY-2409758.